\pdfoutput=1
\documentclass[lettersize,journal]{IEEEtran}
\usepackage{amsmath,amsfonts}
\usepackage{algorithmic}
\usepackage{algorithm}
\usepackage{array}
\usepackage[caption=false,font=normalsize,labelfont=sf,textfont=sf]{subfig}
\usepackage{textcomp}
\usepackage{subcaption}
\usepackage{url}
\usepackage{verbatim}
\usepackage{graphicx}
\usepackage{cite}
\usepackage{booktabs}  

\usepackage{graphicx}
\usepackage{amsmath}
\usepackage[acronym]{glossaries}
\usepackage{tikz} 
\usepackage{adjustbox}
\usepackage{color}
\usepackage{parcolumns}

\newif\iftracked
\trackedfalse    

\iftracked
  \newcommand{\revI}[2]{\noindent\textcolor{blue}{\textbf{[R1.#1]}}~{\color{blue!75}#2}}
  
  \newcommand{\revdel}[1]{\textcolor{gray}{\textbf{[DELETED:]} #1}}
\else
  \newcommand{\revI}[2]{#2}
  
  \newcommand{\revdel}[1]{}
\fi
\usepackage{multicol}
\usepackage{multirow}
\usepackage{rotating}
\usepackage{pdflscape}
\usepackage{afterpage}
\usepackage{float}
\usepackage{amssymb}
\usepackage{amsmath} 

\usepackage{algorithmic}
\usepackage{array}
\usepackage{tabularray}

\definecolor{fondtitre}{RGB}{85,85,85}
\definecolor{fonddeboite}{RGB}{232,232,232}
\definecolor{shadecolor}{RGB}{232,232,232}
\definecolor{melrose}{rgb}{0.8,0.85,1}
\definecolor{Melrose}{rgb}{0.8,0.85,1}
\definecolor{Amber}{rgb}{1,0.756,0.011}
\definecolor{WebOrange}{rgb}{1,0.647,0}
\definecolor{deepblue}{rgb}{0.0, 0.2, 0.6}
\definecolor{blue_swapper}{rgb}{0.0, 0.2, 0.6}

\begin{document}

\newacronym{WCD}{WCD}{Water Column Depth}
\newacronym{LAT}{LAT}{Lowest Astronomical Tide }
\newacronym{HS}{HS}{Hyperspectral}
\newacronym{MS}{MS}{Multispectral}
\newacronym{MSI-2A}{MSI-2A}{MultiSpectral Instrument 2A}
\newacronym{SDB}{SDB}{Satellite Derived Bathymetry}
\newacronym{GUM}{GUM}{Guide to the Expression of Uncertainty in Measurement}
\newacronym{RTE}{RTE}{Radiative Transfer Equation}
\newacronym{BRDF}{BRDF}{ Bi-directional Reflectance Distribution Function}
\newacronym{AOT}{AOT}{Aerosol Optical Thickness}
\newacronym{TOA}{TOA}{Top Of Atmosphere}
\newacronym{BOA}{BOA}{Bottom Of Atmosphere}
\newacronym{IOP}{IOP}{Inherent Optical Properties}
\newacronym{CDOM}{CDOM}{Color Dissolved Organic Matter}
\newacronym{TSM}{TSM}{Total Suspended Matter}
\newacronym{SPM}{SPM}{Suspended Particulate Matter}
\newacronym{LUT}{LUT}{Look-Up Table}
\newacronym{IR}{IR}{InfraRed}
\newacronym{UV}{UV}{Ultra Violet}
\newacronym{NIR}{NIR}{Near InfraRed}
\newacronym{VNIR}{VNIR}{Visible Near InfraRed}
\newacronym{RGB}{RGB}{Red Green Blue}
\newacronym{CMF}{CMF}{Colour Matching Functions}
\newacronym{SA}{SA}{Semi-Analytical}
\newacronym{RT}{RT}{Radiative Transfer}
\newacronym{MNF}{MNF}{Minimum Noise Fraction}
\newacronym{CWBB}{CWBB}{Clear Water With Bright Bedtype}
\newacronym{CWDB}{CWDB}{Complex Water With Dark Bedtype}
\newacronym{TWBB}{TWBB}{Turbid Water With Bright Bedtype}
\newacronym{TWDB}{TWDB}{Turbid Water With Dark Bedtype}

\newacronym{MLE}{MLE}{Maximum Likelihood Estimator}
\newacronym{EFIM}{EFIM}{Expected Fisher Information Matrix}
\newacronym{OFIM}{OFIM}{Observed Fisher Information Matrix}
\newacronym{CI}{CI}{Confidence Intervals}
\newacronym{SNR}{SNR}{Signal to Noise Ratio}

\newacronym{AEU}{AEU}{Average Estimation Update}
\newacronym{APU}{APU}{Average Precision Update}
\newacronym{DAPU}{DAPU}{Directional Average Precision Update}

\title{Quantifying Uncertainty Bounds in Spectral Shallow Water Bathymetry:
 The Profile Likelihood Framework for Semi-Analytical Models}

\author{Mohamed Ali Ghannami,
        Guillaume Sicot,
        Isabelle Quidu,~\IEEEmembership{Member,~IEEE},
        and~Sylvie Daniel,~\IEEEmembership{Senior Member,~IEEE}
\thanks{This work was supported by the Institut France-Qu\'{e}bec Maritime under Grant 208683 and by R\'{e}gion Bretagne.}
\thanks{M. A. Ghannami, G. Sicot, and I. Quidu are with Lab-STICC UMR CNRS 6285, National Institute of Advanced Technologies, ENSTA - Institut Polytechnique de Paris, 29200 Brest, France (e-mail: mohamed-ali.ghannami@ensta-bretagne.fr; guillaume.sicot@ensta.fr; isabelle.quidu@ensta.fr).}
\thanks{S. Daniel is with the Department of Geomatic Sciences, Université Laval, Québec, QC G1V 0A6, Canada (e-mail: sylvie.daniel@scg.ulaval.ca).}
\thanks{Manuscript received November 25, 2025; revised March 24, 2026 and May 5, 2026; accepted September 7, 2026.}}

\markboth{Accepted in IEEE Transactions on Geoscience and Remote Sensing, DOI: 10.1109/TGRS.2026.3736474}%
{Ghannami \MakeLowercase{\textit{et al.}}: Quantifying Uncertainty Bounds in Spectral Shallow Water Bathymetry}

\IEEEpubid{\begin{minipage}{\textwidth}\centering\footnotesize Accepted for publication in IEEE Transactions on Geoscience and Remote Sensing, DOI: 10.1109/TGRS.2026.3736474.\ \copyright~2026 The Authors. This work is licensed under a Creative Commons Attribution 4.0 License (https://creativecommons.org/licenses/by/4.0/).\end{minipage}}

\maketitle   

\begin{abstract}
  Accurate bathymetric mapping in shallow waters is critical
  for coastal monitoring, navigation safety, and environmental
  conservation. While semi-analytical (SA) models have proven
  effective for estimating water column depth (WCD) from optical
  remote sensing data, quantifying the associated uncertainties
  remains a significant challenge. This study introduces
  a unified framework based on profile likelihood to evaluate
  WCD uncertainties in \gls{SA} models, bridging traditional
  variance-covariance methods with likelihood-based inference.
  Through controlled simulations, we systematically assess 
  the performance of first-order statistical tests—likelihood 
  ratio, Wald (observed and expected), and score tests—across
  varying environmental conditions and sensor configurations.

  Our results demonstrate that the expected Fisher information 
  Wald test, equivalent to classical variance-covariance 
  propagation, provides reliable uncertainty bounds in
  hyperspectral settings with clear waters. However, its 
  performance deteriorates in multispectral applications and 
  complex water conditions, where likelihood ratio tests prove 
  more robust. This divergence highlights fundamental limitations 
  in the small-sample regime typical of per-pixel spectral 
  bathymetry. Notably, even under idealized conditions 
  with true parameter initialization, local optimization 
  challenges in \gls{SA} models can compromise both parameter 
  estimation and uncertainty quantification, suggesting even 
  greater challenges in practical applications with less 
  favorable initialization.
  The findings underscore the importance 
  of robust statistical inference in improving the reliability 
  of bathymetric products for operational applications.

  \end{abstract}


\section{Introduction}
Coastal regions are pivotal for human activity, contributing to sustainable development
goals through resource provisioning, employment, and cultural practices~\cite{vazquez,Singh2021}. 
In this setting, accurate bathymetric data is essential for applications like coastal 
monitoring~\cite{Orton2020,Iryanthony2019} and environmental conservation~\cite{Pasquali2021}.
Bathymetric charts are particularly crucial for 
navigation safety. In shallow waters, these data can be acquired through 
various sensing technologies, mainly acoustic and optical sensors~\cite{hydrobook}. 
While acoustic methods have limitations in shallow and hard-to-access zones like
rocky seabeds~\cite{bernardis2023}, optical methods offer a complementary approach.
In the quest for extensive bathymetric data in shallow waters, passive remote sensing with spaceborne 
or airborne optical sensors, has emerged as a particularly effective tool. 
Passive remote sensing provide a cost-effective and scalable solution, especially 
in shallow waters where depth penetrations can range from 5 to 30 meters depending
on sensor performance and environmental factors, mainly water clarity and 
atmospheric conditions \cite{Ashphaq2021}. Passive imagers such as multispectral 
sensors, typically with fewer than a dozen bands, and hyperspectral sensors, 
usually with 100-300 bands, are capable of capturing intricate details in
the visible spectrum.  
In practice, these spectral measurements correlate with \gls{WCD}, 
which serves as a proxy for bathymetry. 
\revI{1/3.1}{In practice, these spectral measurements 
are used to estimate \gls{WCD}, the optical depth 
from the water surface to the seabed, which 
serves as a proxy for bathymetric depth in
 \gls{SDB} applications.}
Building upon Lyzenga's pioneering work~\cite{lyzenga}, empirical models have 
been developed to correlate water surface reflectance with the \gls{WCD}.
 \gls{SA} models, initiated in the works of ~\cite{lee98,lee1999hyperspectral},
go a step further by describing the radiative transfer of light within 
the water column based on a small set of parameters of which \gls{WCD} is part. 
SA models not only estimate \gls{WCD} but also retrieve various biophysical 
properties of the water column, adding layers of information 
without the requirement of in-situ \gls{WCD} measurements. 
\revI{1/3.2}{Since \gls{WCD} is the direct physical
 output of \gls{SA} model inversion --- as opposed 
 to empirical approaches, which regress
  against geodetically-referenced in-situ depth 
  and output bathymetric depth directly --- this paper focuses specifically on \gls{WCD} and its associated uncertainty.}
\IEEEpubidadjcol
Given the wide-ranging applications of bathymetric data, both understanding 
and assessing \gls{WCD} uncertainty are practical necessities. 
Herein, uncertainty of \gls{WCD} derived from passive imagery is
required in order to inform end-users with the derived product quality
and its fitness-for-use. While empirical models for water column depth 
have been refined over the years~\cite{stumpf,Manessa2016,Benshila2022},
these approaches have limitations regarding \gls{WCD} uncertainties. 
\cite{lubac2022} found that the total vertical uncertainty in their 
empirical models was sensitive to a range of factors, including the 
optical conditions of the water and the number of spectral images used.
This suggests that empirical models may not be robust across varying 
datasets and environmental conditions, and may require extensive calibration.
Unlike empirical models, which are often more concerned with predictive
accuracy under specific conditions, \gls{SA} approaches allow for a more 
generalized application and provide a robust framework for both model
evaluation and uncertainty quantification for the \gls{WCD} parameter. 
Only few studies have focused on the uncertainty of biophysical 
parameters in \gls{SA} models, in particular \gls{WCD} 
uncertainty \cite{Garcia2015,jay,jaybottom,1er}. 
In \gls{SA} models, sensor noise and environmental factors introduce a layer of 
complexity that cannot be ignored~\cite{Gao2009,Garcia2015}. Sensor noise, 
which includes factors like photon, dark current, and digitization noise, 
affects the reliability of the retrieved parameters. Environmental factors
such as atmospheric aerosol and sunglint effects in spectral imagery add another 
layer of uncertainty. Moreover, the precision in retrieved \gls{WCD} and benthic
type varies with factors like water turbidity and substrate albedo, making
the quantification of uncertainty a pixel-dependent 
task~\cite{hedley,Garcia2015,mckinna2019approach}.

This article investigates the fundamental limitations in 
uncertainty quantification for remotely sensed shallow water \gls{WCD} 
through SA models. We examine how the profile likelihood framework unifies
traditional variance-covariance approaches with likelihood-based 
inference, particularly in the context of single-pixel analyses where
sample sizes are inherently limited. Building on established Maximum Likelihood Estimator (\gls{MLE}) methods
 \cite{jay, jaybottom, guillaume2023determination}, we demonstrate how different statistical tests - from classical Wald tests 
to likelihood ratio statistics - perform under various environmental and sensor configurations.
To rigorously evaluate these statistical approaches, we adopt a controlled 
simulation framework where true parameters are known. This experimental design is critical for isolating the inherent statistical
limitations from practical complications such as 
residual biases related to atmospheric effects and 
sensor calibration. 

The main contributions of this paper are threefold:

\begin{enumerate}
  \item Unification of variance-covariance and likelihood-based approaches through the profile likelihood framework, revealing their theoretical connections for evaluating \gls{WCD} uncertainties.
  \item Systematic characterization of statistical test performance across water conditions and spectral resolutions, demonstrating fundamental limitations in \gls{WCD} inference.
  \item Demonstration of how local optimization challenges in \gls{SA} models affect both parameter estimation and uncertainty quantification, even under idealized conditions.
\end{enumerate}

The remaining structure of this paper is organized as follows: Section II reviews 
related work in radiometric uncertainty evaluation and \gls{WCD} estimation. 
Section III presents the theoretical framework, including the \gls{SA} model 
formulation and likelihood-based inference methodology. Section IV describes 
our experimental setup and details the simulation framework. Section V presents
the results of our \gls{WCD} profiling analysis and evaluates the performance of 
different statistical tests. Section VI discusses the implications of our findings
for practical applications and compares hyperspectral and multispectral performance. 
Finally, Section VII concludes the paper and outlines future research directions.

\section{Related Works}

The uncertainty quantification in \gls{WCD} estimation through
\gls{SA} models necessitates understanding two distinct but 
interconnected sources of uncertainty: those associated with the
input remote sensing reflectance (\(R_{rs}\)) and those stemming 
from the model parameter estimation process. The remote sensing 
reflectance serves as the primary input to these models, with its
uncertainties directly influencing the reliability of retrieved 
environmental parameters, particularly \gls{WCD}. We first examine \(R_{rs}\)
uncertainties and their evaluation before addressing
parameter uncertainty estimation in \gls{SA} models.

\subsection{\(R_{rs}\) Uncertainty Sources}

Remote sensing reflectance estimation faces challenges from both instrumental
and environmental factors, each contributing significantly to the overall 
uncertainty budget \cite{Gao2009,Garcia2015}. At the instrumental level, 
sensor-specific uncertainties arise from multiple sources: photon noise,
which follows Poisson statistics and varies with signal intensity; 
dark current noise, inherent to the detector electronics; and 
digitization noise from the analog-to-digital conversion process. 
Environmental factors introduce additional complexity through 
atmospheric effects, sunglint contamination and adjacency effects, particularly 
challenging in coastal waters.

Methods for quantifying uncertainties in \(R_{rs}\) can be categorized
into three main approaches: empirical, scene-based, and pixel-based techniques. 
Empirical methods, as demonstrated by \cite{Hu2013} and \cite{melin2015}, 
rely on direct comparisons between satellite observations and in-situ 
measurements. These studies consistently reveal that atmospheric
correction uncertainties typically dominate over sensor calibration effects,
particularly in complex coastal environments. While valuable for
understanding general sensor performance, these methods are 
limited in their ability to capture pixel-specific uncertainties 
crucial for bathymetric applications.

Scene-based approaches introduced the concept of Noise Equivalent 
Reflectance Difference (NE\(\Delta\)R) \cite{Gordon1997,Wettle2004,Sagar2014},
providing a more sophisticated framework for uncertainty quantification. 
This metric incorporates both sensor noise and environmental factors 
without requiring detailed sensor characterization.
However, its effectiveness diminishes in optically shallow
waters where environmental variability can be extreme 
\cite{hedley,jay,jaybottom,Garcia2015,Garcia2014}, 
limiting its applicability in bathymetric contexts.

More recently, pixel-based methods have emerged
\cite{Jia2018,nicolas,pachniak2023quantifying}, offering more 
detailed approaches to uncertainty quantification. 
These methods propagate sensor uncertainties through both 
atmospheric and sunglint correction processes at the individual pixel level.
Despite their sophistication, they face limitations from simplified assumptions
in aerosol models and surface characterization, particularly in the treatment 
of the \gls{BRDF}. Even under optimal conditions, these methods indicate 
residual uncertainties in \(R_{rs}\) values ranging from 30\% to 50\%. 
Such uncertainties become particularly critical in shallow water applications,
where the non-linear nature of radiative transfer can amplify their effects
on \gls{WCD} estimation. This sensitivity is especially pronounced in areas 
where small variations in \(R_{rs}\) can lead to significant changes in 
derived \gls{WCD} values due to the complex interaction between water column
properties and bottom reflectance.

While these \(R_{rs}\)  uncertainty quantification approaches provide
 valuable insights into measurement reliability, their application to \gls{SA} 
bathymetric models presents unique challenges. Whereas empirical models 
directly relate reflectance to \gls{WCD} through statistical relationships, 
\gls{SA} models explicitly incorporate the physical processes of
light propagation through the water column. This physics-based approach 
means that \(R_{rs}\)  uncertainties propagate through complex radiative transfer equations,
potentially amplifying their effects on derived parameters in non-linear ways.
Moreover, the interdependence between water column properties, bottom 
characteristics, and \gls{WCD} in \gls{SA} models means that \(R_{rs}\)  
uncertainties can simultaneously affect multiple parameter estimates,
creating correlated uncertainty structures that simpler propagation 
methods may not adequately capture. These considerations highlight the 
need for sophisticated uncertainty quantification frameworks specifically 
designed for the parameter estimation challenges inherent in 
\gls{SA} bathymetric modeling.
\subsection{\gls{WCD} Uncertainty in Semi-Analytical Models}

The quantification of uncertainties in \gls{SA} models presents distinct
 methodological challenges that have been approached through various 
 statistical frameworks. Among these, bootstrapping techniques 
\cite{hedley2010propagating, Garcia2015} have been widely employed 
to characterize parameter uncertainties through empirical distribution
estimation. While conceptually straightforward in its resampling approach,
bootstrapping faces significant limitations in the context of \gls{SA} models.
Beyond the obvious computational burden of repeated inversions that makes it 
impractical for large-scale applications and real-time processing, 
bootstrapping encounters a more fundamental challenge: the reliability
of parameter estimation at each resampling iteration. This becomes 
particularly critical in scenarios with limited statistical information
such as low signal quality, increased depths, or complex water properties
\cite{petit,Garcia2015}, where optimization procedures frequently fail 
to converge to meaningful solutions.

Classical error propagation methods, particularly the variance-covariance 
approaches prescribed by measurement standards Guide to Expression of
Uncertainty in Measurement (\gls{GUM}) in \cite{GUM}, have seen limited 
application in shallow water \gls{SA} models. \cite{Garcia2015} specifically 
highlighted the inherent difficulties of extending such frameworks to shallow 
water environments, where model complexity significantly complicates uncertainty
propagation. While these variance-covariance methods have found success in 
ocean color applications 
\cite{lee2010uncertainties,maritorena2004consistent,salama2011ensemble} 
- where models are simplified by excluding \gls{WCD} and seabed parameters -
even these relatively simpler contexts reveal significant limitations. 
Studies by \cite{Sydor2003} and \cite{DefoinPlatel2007} demonstrated 
that ocean color models suffer from ill-posedness and ambiguity issues 
that fundamentally impact uncertainty evaluation through variance-covariance
frameworks. These challenges become even more pronounced in shallow water
applications where the addition of \gls{WCD} and seafloor parameters 
substantially increases model complexity and the potential for
multiple solutions.

The emergence of hyperspectral sensors, offering hundreds of spectral bands, 
has transformed the capabilities of parameter estimation in \gls{SA} models. 
Studies have consistently demonstrated improved \gls{WCD} estimation accuracy 
with hyperspectral measurements compared to traditional multispectral data
\cite{Garcia2015, Garcia2018}. \cite{sicot} provided crucial insights 
through their analysis of parameter estimation using global optimization
techniques. Their work demonstrated that hyperspectral data enables 
reliable estimation of not just \gls{WCD} but all model parameters 
- including water composition and seafloor properties. This 
confirms that estimation challenges arise primarily from measurement 
noise and model complexity rather than any inherent limitations in the
estimation problem itself. 

These findings point to a fundamental challenge in parameter 
estimation for spectral bathymetry. While theoretical analyses 
suggest that parameters could be perfectly recovered under ideal
conditions, real-world measurements introduce noise and uncertainties 
that complicate the estimation process \cite{Wieland2021,pronzato,Lam2022}.
This distinction becomes particularly critical when working with limited 
spectral data, where uncertainty ranges may exhibit asymmetric behavior 
- that is, the confidence in parameter values may be significantly different
above and below the estimated value \cite{1er}. The challenge intensifies in 
satellite-based \gls{WCD} estimation, where we must simultaneously 
determine multiple environmental parameters such as water composition,
seafloor properties, and \gls{WCD} itself, all from a limited set of 
spectral measurements. This interplay between limited spectral measurements 
and multiple unknown parameters creates a complex uncertainty landscape that
requires careful consideration in operational settings, particularly where traditional 
statistical assumptions about sample size and parameter identifiability may not hold.

Recent developments have focused on likelihood-based methods, 
offering a promising framework for addressing these challenges. 
\cite{jay} developed analytical approaches for tracking uncertainties
in \gls{SA} models, while \cite{guillaume2023determination} explored
ways to incorporate prior knowledge about reasonable parameter values.
However, specifying appropriate constraints remains challenging in dynamic
coastal environments where conditions change frequently. Our work builds 
on these advances by proposing a unified framework that systematically 
evaluates different uncertainty 
quantification methods for the water column depth, providing practical 
guidance for both hyperspectral and multispectral applications 
across varying environmental conditions.

\section{Methodology}

In this study, \gls{WCD} is modeled through the \gls{SA} approach proposed by Lee's model for describing the remote sensing 
reflectance in terms of biophysical parameters \cite{lee1999hyperspectral}. 
The mathematical inference framework presented in this study involves
extensive notation as summarized in Table \ref{tab:inference_notations}.

\subsection{Theoretical framework}
For a given pixel
in a spectral data cube, the \( R_{rs} \) data denoted $\textbf{y}$ is 
used to derive the parameter vector which is represented as \( \boldsymbol{\theta} = (\psi, \boldsymbol{\omega}) \), 
where \( \psi \) is the parameter of interest, \textit{i.e.} : \gls{WCD}. 
The vector \( \boldsymbol{\omega} \) encapsulates various nuisance parameters and can be decomposed into
\( \boldsymbol{\omega} = (\boldsymbol{\omega}_c, \boldsymbol{\omega}_b, \boldsymbol{\omega}_g, \boldsymbol{\omega}_e) \). 
Here, \( \boldsymbol{\omega}_c \) and \( \boldsymbol{\omega}_b \) are associated with the water \gls{IOP}s and bottom properties, respectively. 
The viewing geometry is captured by \( \boldsymbol{\omega}_g \), which itself is decomposed into \( \boldsymbol{\omega}_g = (\xi_s, \xi_v) \), 
representing the solar zenith and viewing angles. \( \boldsymbol{\omega}_g \) is considered fixed as it can be extracted directly from sensor 
metadata, assuming the sensor's geometric performance is reliable. For our study, we selected a solar zenith angle of $30^\circ$~and a 
viewing angle of $0^\circ$, with the latter representing the nadir direction for off-zenith angles, 
typically corresponding to standard surveying geometries.
The vector \( \boldsymbol{\omega}_e \) comprises empirical coefficients from the \gls{SA} model, retrieved from
literature, and will be denoted by \( \gamma \), \( \beta \), or \( \nu \), indexed according to their specific use as described in the following. 
These empirical parameters, characterized by regional and seasonal variability, 
will be considered as fixed, thus reducing the nuisance parameters to $\boldsymbol{\omega} = (\boldsymbol{\omega}_c, \boldsymbol{\omega}_b)$.
This approach is common in \gls{SA} studies, 
as these parameters are typically derived from established methods 
such as laboratory analyses of water samples or data-driven procedures. 
Considering them as fixed allows models to be applied in standard 
configurations, involving literature-derived empirical coefficients
and spectral parameters rather than optimizing them from the data.

In order to derive the uncertainties of the interest parameter
akin to the \gls{WCD} in this paper,
we use the profile likelihood and its associated first-order test statistics.
The profile likelihood of the parameter of interest $\psi$ is 
defined as:
\begin{equation}
    \ell_p(\psi; \textbf{y}) = \max_{\boldsymbol{\omega}} \ell(\psi, \boldsymbol{\omega}; \textbf{y}) = \ell(\psi, \hat{\boldsymbol{\omega}}_\psi; \textbf{y})
\end{equation}

\begin{figure}[ht]
  \hspace{-1cm}
    \centering  
    \includegraphics[scale=0.28]{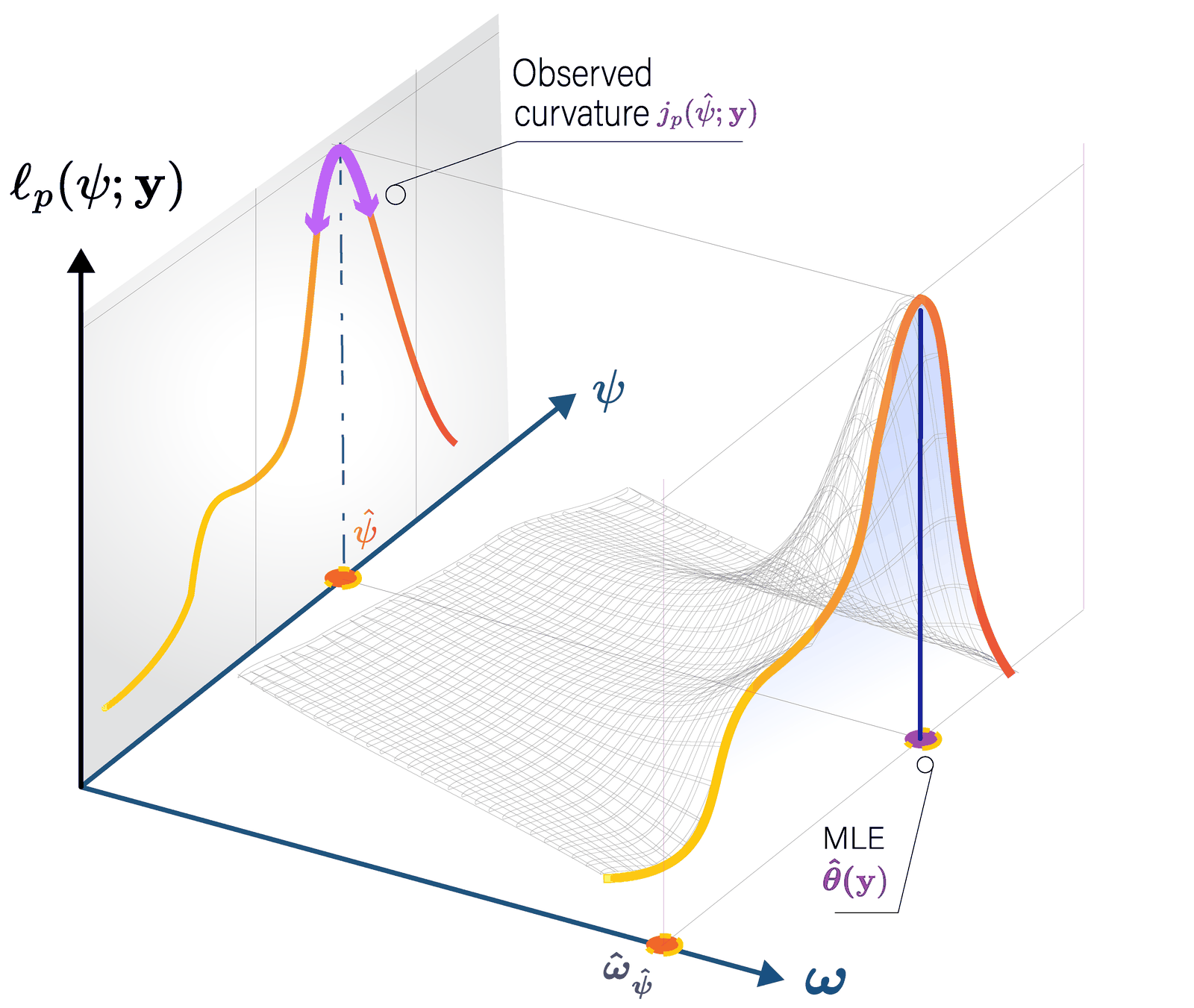}
    \caption[Geometric illustration of profile Likelihood]{Geometric illustration of profile
     Likelihood surface for a two-dimensional parameter space.}
     \label{fig:lp}
\end{figure}

Where $\ell(\psi, \boldsymbol{\omega}; \textbf{y})$ is the log-likelihood function of the 
parameters $\psi$ and $\boldsymbol{\omega}$ and $\hat{\boldsymbol{\omega}}_\psi$ is the constrained \gls{MLE}
of the nuisance parameters for each value of $\psi$. In essence, the profile likelihood $\ell_p(\psi; \textbf{y})$ 
simplifies the likelihood function by focusing on $\psi$ and 
adjusting the nuisance parameters $\boldsymbol{\omega}$ accordingly. 
This is achieved by maximizing the likelihood with respect 
to the nuisance parameter, resulting in the constrained 
constrained \gls{MLE} $\boldsymbol{\hat\omega}_\psi$ for each
value of $\psi$ (see Fig. \ref{fig:lp}). 

For a scalar parameter
of interest \( \psi \), first-order inference can be 
performed based on the following test statistics \cite{brazzal}:

\begin{align}
r(\psi;\textbf{y}) &= \text{sign}(\hat{\psi} - \psi) \sqrt{2 (\ell_p(\hat{\psi};\textbf{y}) - \ell(\psi;\textbf{y}))} \label{eq:r}, \\
t_o(\psi;\textbf{y}) &= j_p(\hat{\psi};\textbf{y})^{1/2} (\hat{\psi} - \psi) \label{eq:t_o}, \\
t_e(\psi;\textbf{y}) &= i_p(\hat{\psi})^{1/2} (\hat{\psi} - \psi) \label{eq:t_e}, \\
s(\psi;\textbf{y}) &= j_p(\hat{\psi};\textbf{y})^{-1/2} \frac{\partial \ell_p(\psi;\textbf{y})}{\partial \psi} \label{eq:s}
\end{align}

Here, \( r \) is the signed likelihood ratio statistic, \( t_o \) and \( t_e \) are the Wald test
statistics using observed and expected Fisher information respectively, and \( s \) is the score statistic.
\( j_p(\hat{\psi};\textbf{y}) \) and \( i_p(\hat{\psi}) \) are the observed and expected 
Fisher information of \( \psi \) at the \gls{MLE} $\hat{\psi}$, given by:

\begin{align}
j_p(\hat{\psi};\textbf{y}) &= -\frac{\partial^2 \ell_p(\hat{\psi};\textbf{y})}{\partial \psi^2}, \\
i_p(\hat{\psi}) &= \mathbb{E}[j_p(\hat{\psi};\textbf{y})]
\end{align}

Under certain regularity conditions, the main results of first-order inference can be summarized as follows:
\begin{equation} 
  st(\tilde{\psi};\textbf{y}) \xrightarrow{O(n^{-1/2})} \mathcal{N}(0, 1)
  {\label{eq:stat_tests_standard}}
  \end{equation}
\noindent where \( st \in \{r, t_o, t_e, s\} \), and \( n \) is the number of independent samples.
  
This stipulates that under the null hypothesis {(\textit{i.e.} at 
the true value \(\tilde\psi\))}, all these test
statistics asymptotically follow a standard normal 
distribution. In this paper, we focus on the 
evaluation of coverage probability to assess the reliability of these intervals. 
Coverage probability is defined as the frequency with which the true value
of the parameter of interest, \( \psi \), falls within the estimated 
\gls{CI} \( \mathcal{A}(\textbf{y}) \). Mathematically, 
coverage probability is expressed as:

\begin{equation}
P(\tilde\psi \in \mathcal{A}(\textbf{y})) = 1-\alpha
\end{equation}
Here, \( \alpha \) represents the significance level, commonly set at 0.05 equivalent to 95\% confidence level.

To construct these \gls{CI}, we employ the first-order test statistics based on (\ref{eq:stat_tests_standard}) as follows:

\begin{align}
\text{For } r: \quad & \psi \in \{ \psi: |r(\psi;\mathbf{y})| \leq z_{\alpha/2} \} \\
\text{For } t_o: \quad & \psi \in \left\{ \hat\psi \pm z_{\alpha/2} \sqrt{j_p(\psi;\mathbf{y})} \right\} \\
\text{For } t_e: \quad & \psi \in \left\{ \hat\psi \pm z_{\alpha/2} \sqrt{i_p(\psi)} \right\} \\
\text{For } s: \quad & \psi \in \{ \psi: |s(\psi;\mathbf{y})| \leq z_{\alpha/2} \}
\end{align}

Since the interest parameter is considered scalar, \( z_{\alpha/2} \) 
is the critical value corresponding to the \( \alpha/2 \) 
tail of a standard normal distribution,
generally approximated as 1.96 for a 95\% confidence level.

\begin{figure}[ht]
\centering  
\includegraphics[scale=0.3]{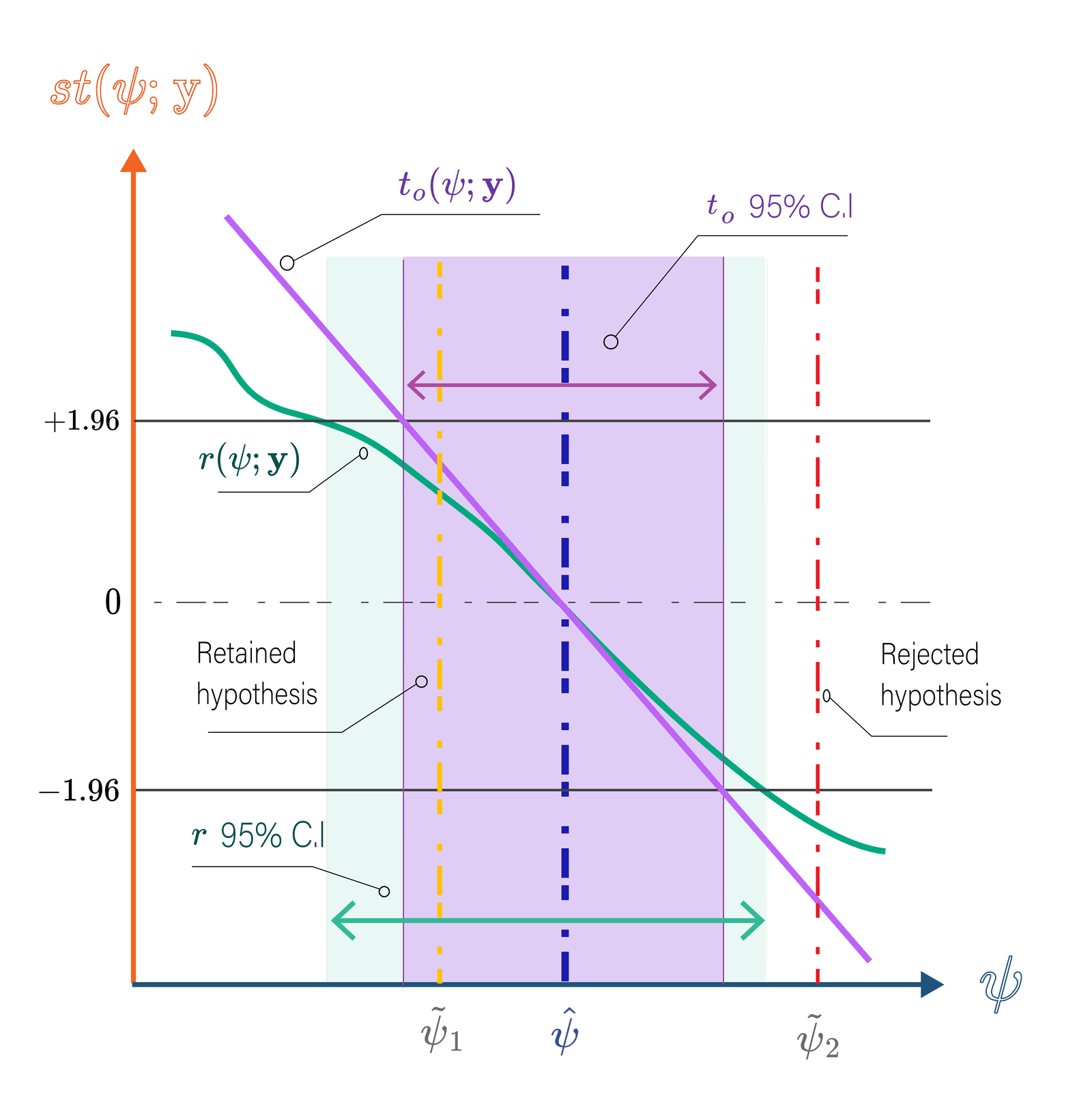}
\caption[First-order statistical tests and hypothesis testing]{First-order
 statistical tests and hypothesis testing: 
The red  (resp. yellow) line indicates a case of  (resp. non) rejection of the null hypothesis $\tilde\psi$ (Type I error)
which should theoretically occur 5\% of the time on the long run,
depending on the statistical test performance. The \( r \) curve, depicted for a small sample scenario,
is expected to exhibit a
linearly decreasing trend analogous to Wald tests with increasing sample size.}
\label{fig:st}
\end{figure}

In order to empirically validate the coverage probability of \gls{CI}, 
we use Monte Carlo simulations to generate sample data 
$\textbf{y}$ from the actual parameter values $\boldsymbol{\tilde\theta}$.
Specifically, we calculate the observed test statistic {for each sample $\textbf{y}$} at the true
 \( \tilde{\psi} \) value under the null hypothesis.
An observed test statistic's absolute value greater than 
\( z_{\alpha/2} = 1.96 \) would indicate that the null
hypothesis can be rejected at the \( 95\% \) confidence level. As highlighted
in Fig. \ref{fig:st}, 
Wald tests $t_o$ and $t_e$ are linear 
slopes by construction, which yield symmetric \gls{CI} around the \gls{MLE} $\hat\psi$. 
On the other hand, $r$ and $s$ effectively capture the
asymmetric behavior which can arise in non linear setups.

In the realm of first-order inference for \gls{WCD} analysis 
in remote sensing, especially in radiometric and geometric contexts, 
the use of profile likelihood remains notably scarce, 
despite its longstanding presence as a statistical tool. It's important to 
highlight that the use
of the expected Fisher information Wald test under Gaussian errors is 
effectively equivalent to variance-covariance
propagation \ref{app:var_covar_equiv_mle}. Furthermore, \( j_p \) (resp. \( i_p \)) are 
equivalent to the diagonal elements at \( \psi \) blocks
of the observed (resp. expected) Fisher information matrices traditionally
computed in multidimensional inference on \( \boldsymbol{\theta} \), 
the full parameter set. The scalar $\psi$ test statistics are in harmony with 
established multidimensional inference techniques. 
In the context of a multidimensional parameter, the test statistics
converge to a chi-squared distribution, where the 
degrees of freedom match the dimensionality of the
interest parameter. This ensures that focusing on
a scalar parameter \( \psi \) does not deviate 
from traditional multidimensional inference methods. These equivalences subtly implies 
an implicit use of profile likelihood concepts through Wald tests,
making the profile likelihood framework more 
extensive with the likelihood ratio and the score as 
additional tools as well as applicable for quantifying
uncertainties. Consequently, this framework can be comprehensively 
placed under the umbrella
of profile likelihood, underscoring its 
broader relevance and untapped potential for uncertainties analysis in remote sensing.

\begin{table}[t]
\centering
\footnotesize
\caption{Statistical Inference Notations}
\label{tab:inference_notations}
\begin{tblr}{
  width = 0.95\linewidth,
  colspec = {Q[110]Q[270]},
  cells = {c},
  row{1} = {melrose},
  row{odd} = {melrose},
  hline{1,2} = {-}{0.08em},
  hline{Z} = {-}{0.08em},
}
\textbf{Symbol} & \textbf{Description} \\
$\boldsymbol{\tilde\theta}$ & True parameter values \\
$\textbf{y}$ & Observed data \\
$\boldsymbol{\theta}$ & Model parameters \\
$\psi$ & Interest parameter \\
$\boldsymbol{\omega}$ & Nuisance parameters \\
$L(\psi, \boldsymbol{\omega}; \textbf{y})$ & Likelihood function \\
$\ell(\psi, \boldsymbol{\omega}; \textbf{y})$ & Log-likelihood function \\
$\boldsymbol{\hat\theta}$ & Maximum Likelihood Estimator \\
$\hat{\psi}$ & MLE of interest parameter \\
$\hat{\boldsymbol{\theta}}_\psi$ & Constrained MLE \\
$\hat{\boldsymbol{\omega}}$ & MLE of nuisance parameters \\
$\ell_p(\psi;\textbf{y})$ & Profile log-likelihood \\
$r(\psi;\textbf{y})$ & Likelihood ratio statistic \\
$t_o(\psi;\textbf{y})$ & Wald test (Observed Information) \\
$t_e(\psi;\textbf{y})$ & Wald test (Expected Information) \\
$s(\psi;\textbf{y})$ & Score statistic \\
$j_p(\hat{\psi};\textbf{y})$ & Observed Fisher Information of $\psi$ \\
$i_p(\hat{\psi})$ & Expected Fisher Information of $\psi$ \\
$\boldsymbol{j}(\boldsymbol{\hat\theta};\textbf{y})$ & Observed Fisher Information Matrix \\
$\boldsymbol{i}(\boldsymbol{\hat\theta})$ & Expected Fisher Information Matrix \\
$\mathcal{A}(\textbf{y})$ & Confidence interval \\
$\alpha$ & Significance level \\
$z_{\alpha/2}$ & Critical value \\
\end{tblr}
\end{table}

\begin{table}[t]
\centering
\footnotesize
\caption{Semi-Analytical Model Notations}
\label{tab:radiometric_notations}
\begin{tblr}{
  width = 0.95\linewidth,
  colspec = {Q[110]Q[270]},
  cells = {c},
  row{1} = {melrose},
  row{odd} = {melrose},
  hline{1,2} = {-}{0.08em},
  hline{Z} = {-}{0.08em},
}
\textbf{Symbol} & \textbf{Description} \\
$\gamma, \nu, \beta$ & Semi-analytical coefficients \\
$\boldsymbol{R}_{rs}^+$ & Above-surface reflectance (sr$^{-1}$) \\
$\boldsymbol{R}_{rs}^-$ & Subsurface reflectance (sr$^{-1}$) \\
$z$ & Water Column Depth (m) \\
$\boldsymbol{\omega}_c$ & Water column parameters \\
$\boldsymbol{\omega}_b$ & Bottom parameters \\
$\boldsymbol{\rho}(\boldsymbol{\omega}_b)$ & Bottom reflectance \\
$\alpha_m$ & Mixing ratio \\
$\boldsymbol{a}$ & Absorption coefficient (m$^{-1}$) \\
$\boldsymbol{b_b}$ & Backscattering coefficient (m$^{-1}$) \\
$\boldsymbol{a}_\text{w}$ & Water absorption (m$^{-1}$) \\
$\boldsymbol{a}_\phi$ & Phytoplankton absorption (m$^{-1}$) \\
$\boldsymbol{a}_\text{cdom}$ & \gls{CDOM} absorption (m$^{-1}$) \\
$\boldsymbol{a}_\text{nap}$ & Non-algal absorption (m$^{-1}$) \\
$C_{\text{chl}}$ & Chlorophyll-A concentration (mg m$^{-3}$) \\
$C_{\text{cdom}}$ & \gls{CDOM} absorption coefficient (m$^{-1}$) \\
$C_{\text{tsm}}$ & \gls{TSM} concentration (g m$^{-3}$) \\
\end{tblr}
\end{table}

\subsection{Shallow water Semi-Analytical model {\label{sec:lee_model}}}

The above-surface remote sensing reflectance \( \boldsymbol{R}_{rs}^+\) in terms of \gls{WCD} \( z \), water column 
parameters \( \boldsymbol{\omega}_c \), bottom parameters \( \boldsymbol{\omega}_b \) is related to the subsurface 
remote sensing reflectance \( \boldsymbol{R}_{rs}^-\) by:
\begin{equation}
  \boldsymbol{R}_{rs}^+(z, \boldsymbol{\omega}_c, \boldsymbol{\omega}_b) = \frac{\gamma_{s} \cdot \boldsymbol{R}_{rs}^-(z, \boldsymbol{\omega}_c, \boldsymbol{\omega}_b)}{1 - \beta_{s} \cdot \boldsymbol{R}_{rs}^-(z, \boldsymbol{\omega}_c, \boldsymbol{\omega}_b)}
\end{equation}

The mathematical framework for the semi-analytical
model adopted in this study involves
the notation presented in Table \ref{tab:radiometric_notations}. While this model is underpinned by radiative transfer theory, 
it incorporates empirical components such as bottom spectra and chlorophyll 
absorption spectra, which are inherently variable and region-specific. These 
empirical components are selected based on established literature values, recognizing 
their variability \cite{Pope97,bricaud_chl,bricaud,bpglobal,lee1999hyperspectral,Morel}. The proposed approach allows to focus 
on the primary objectives of the model without the need for detailed justification
of these empirical choices, which, although secondary, are essential for the model's application.

The empirical coefficients \( \gamma_{s} \) and \( \beta_{s} \)  encapsulate the surface effects of factors such as
viewing geometry, water properties and internal reflections. These are fixed at the values 
\( \gamma_{s} = 0.5\) and  \( \beta_{s} = 1.5\) as provided in \cite{lee1999hyperspectral}.

The subsurface remote sensing reflectance \( \boldsymbol{R}_{rs}^- \) is modeled as the sum of the
remote sensing reflectances due to water column \(  \boldsymbol{R}_{rs}^C  \) and bottom \(  \boldsymbol{R}_{rs}^B \):

\begin{equation}
  \boldsymbol{R}_{rs}^-(z, \boldsymbol{\omega}_c, \boldsymbol{\omega}_b) = \boldsymbol{R}_{rs}^C (z, \boldsymbol{\omega}_c) + \boldsymbol{R}_{rs}^B (z, \boldsymbol{\omega}_c, \boldsymbol{\omega}_b)
\end{equation}

The remote sensing reflectance due to bottom contribution \( \boldsymbol{R}_{rs}^B \) is expressed as:

\begin{equation}
  \boldsymbol{R}_{rs}^B(z,\boldsymbol{\omega}_c,\boldsymbol{\omega}_b) = \frac{\boldsymbol{\rho}(\boldsymbol{\omega}_b)}{\pi} \exp(-\boldsymbol{K}^B_u(\boldsymbol{\omega}_c) \cdot z)
\end{equation}

Here, \( \boldsymbol{\rho}(\boldsymbol{\omega}_b) \) is the seabed reflectance, which is modeled as a 
sum-to-one linear mix of sand and vegetation reflectances. We use two empirical spectra, \( \boldsymbol{\rho}_1 \) 
and \( \boldsymbol{\rho}_2 \), representing the spectra for sand and vegetation, respectively. The parameter \( \alpha_m \) 
controls the mixing ratio between them as the sole component of \( \boldsymbol{\omega}_b \) :

\begin{equation}
  \boldsymbol{\rho}(\alpha_m) = (1-\alpha_m) \boldsymbol{\rho}_1 + \alpha_m \boldsymbol{\rho}_2
\end{equation}

The remote sensing reflectance due to water column \( \boldsymbol{R}_{rs}^C \) is given by:

\begin{equation}
  \boldsymbol{R}_{rs}^C(z,\boldsymbol{\omega}_c) = \boldsymbol{r}_{rs}^{dp}(\boldsymbol{\omega}_c) \left(1 - \exp(-\boldsymbol{K}^C_u(\boldsymbol{\omega}_c) \cdot z)\right)
\end{equation}

\( \boldsymbol{K}^C_u \) and \( \boldsymbol{K}^B_u \) 
are the diffuse attenuation coefficients for the water column and bottom, respectively, whereas $\boldsymbol{r}_{rs}^{dp}$ is the remote-sensing reflectance for optically
deep waters. These are functions of the optically active constituents of the water, all parameterized by \( \boldsymbol{\omega}_c \) : 

\begin{align}
\boldsymbol{K}_b(\boldsymbol{\omega}_c) &= \left( \frac{1}{\cos(\xi_s)} + \frac{\gamma_b \sqrt{1 + \beta_b \boldsymbol{u}(\boldsymbol{\omega}_c)}}{cos(\xi_v)} \right) \boldsymbol{\kappa} (\boldsymbol{\omega}_c), \\
\boldsymbol{K}_c(\boldsymbol{\omega}_c) &= \left( \frac{1}{\cos(\xi_s)} + \frac{\gamma_c \sqrt{1 + \beta_c \boldsymbol{u}(\boldsymbol{\omega}_c)}}{cos(\xi_v)} \right) \boldsymbol{\kappa} (\boldsymbol{\omega}_c), \\
\boldsymbol{r}_{rs}^{dp}(\boldsymbol{\omega}_c) &= (\gamma_{dp} + \beta_{dp} \boldsymbol{u}(\boldsymbol{\omega}_c)) \boldsymbol{u}(\boldsymbol{\omega}_c).
\end{align}

The empirical coefficients are set to the values $\gamma_c  = 1.03, \gamma_b = 1.04, \beta_c = 2.4, \beta_b = 5.4$
 as provided in \cite{lee1999hyperspectral}.
Here, \( \boldsymbol{\kappa} \) and \( \boldsymbol{u} \) are defined as:

\begin{equation}
\begin{aligned}
  \boldsymbol{\kappa} (\boldsymbol{\omega}_c) &= \boldsymbol{a}(\boldsymbol{\omega}_c) + \boldsymbol{b_b}(\boldsymbol{\omega}_c), \quad
  \boldsymbol{u}(\boldsymbol{\omega}_c) &= \frac{\boldsymbol{b_b}(\boldsymbol{\omega}_c)}{\boldsymbol{\kappa} (\boldsymbol{\omega}_c)}.
\end{aligned}
\end{equation}

For modeling the absorption \( \boldsymbol{a} \) and backscattering \( \boldsymbol{b_b} \) coefficients, we consider
the absorption contributions from water, phytoplankton, Colored Dissolved Matter (\gls{CDOM}), and Total Suspended Matter (\gls{TSM}), as well as the backscattering contributions from water and \gls{TSM}. For a detailed description of the
equations modeling absorptions and backscattering, the reader is referred to Section \ref{app:iops}.

Consequently, the primary parameters besides the \gls{WCD} (\textit{i.e.} \( z \)), are the seabed albedo mixing ratio \( \alpha_m \),  
and $C_\text{chl}$ the concentration of chlorophyll (in \( \text{mg m}^{-3} \)), $C_\text{cdom}$ the absorption coefficient of \gls{CDOM} (in \( \text{m}^{-1} \)), 
and  $C_\text{tsm}$ the concentration of \gls{TSM} (in \( \text{g m}^{-3} \)).  
Formally, $\boldsymbol{\theta} = \{\psi: z, \boldsymbol{\omega} : \{\boldsymbol{\omega_b}: \alpha_m,\boldsymbol{\omega_c} : \{C_\text{Chl},C_\text{CDOM},C_\text{TSM}\}\}\}$.
To ensure unconstrained optimization while maintaining physical parameter constraints, we 
employed reparameterization strategies for all nuisance parameters (\textit{i.e.}\ estimated 
parameters except \gls{WCD}). Concentration parameters were log-transformed to ensure positivity:
 $C_\text{chl} = \exp(\theta_\text{chl})$, $C_\text{cdom} = \exp(\theta_\text{cdom})$, $C_\text{tsm} = \exp(\theta_\text{tsm})$, where $\theta_c \in \mathbb{R}$. 
 The bottom albedo mixing coefficient used the logistic sigmoid: $\alpha_m = 1/(1 + \exp(-\theta_{\alpha_m}))$, $\theta_{\alpha_m} \in \mathbb{R}$, 
 mapping $\theta_{\alpha_m} \to -\infty$ to $\alpha_m \to 0$ (vegetation) and $\theta_{\alpha_m} \to +\infty$ to $\alpha_m \to 1$ (sand). 
 These transformations allow optimization over the entire real domain while automatically preserving physical constraints, satisfying the
  regularity conditions required for asymptotic statistical inference.

To account for sensor-specific effects and the integration over the sensor's 
spectral response functions, we redefine the sensor-dependent above-surface 
remote sensing reflectance, denoted as \( \boldsymbol{R}_{rs}^{S+} \). 
This measurement includes the dependency on each of the sensor's discrete
sampling bands. The integrated remote sensing reflectance for
each band \( b_k \) is defined as:
\begin{equation}
\boldsymbol{R}_{rs}^{S+}(z, \boldsymbol{\omega}, b_k) = 
\frac{\int_{\lambda} S_k(\lambda) \cdot
\boldsymbol{R}_{rs}^+(z, \boldsymbol{\omega}, \lambda) \, 
d\lambda}{\int_{\lambda} S_k(\lambda) \, d\lambda}
\end{equation}
where \( b_k \) represents each discrete spectral band of the sensor,
and \( S_k(\lambda) \) is the spectral response function of the sensor
at each wavelength \( \lambda \) within the band \( b_k \). 
This normalized integration ensures that the reflectance values are
adjusted for the overall sensitivity of the sensor across its bandwidth.

\subsection{Statistical modeling}

In our model, we consider a Gaussian measurement model for the remote sensing reflectance at \( m \) 
different wavelengths, spanning the \gls{VNIR} spectrum \cite{jay}.
For each band \( b_k \), \( k = 1, \ldots, m \), the observed above-surface remote
sensing reflectance \( \textbf{y}(b_k) \) is modeled as:
\begin{equation}
\textbf{y}(b_k) = \boldsymbol{R}_{rs}^{S+}(z, \boldsymbol{\omega}, b_k) + \boldsymbol{\epsilon}(b_k)
\end{equation}
Here, \(  \boldsymbol{R}_{rs}^{S+}(z, \boldsymbol{\omega}, b_k) \) is the true above-surface remote sensing reflectance at \gls{WCD} \( z \), 
with parameter vector \( \boldsymbol{\omega} \), and at band \( b_k \). The term \( \boldsymbol{\epsilon}(b_k) \) represents the 
measurement noise at band \( b_k \), \( \boldsymbol{\epsilon} \) is assumed to follow a multivariate Gaussian distribution
$\boldsymbol{\epsilon} \sim \mathcal{N}(0, \boldsymbol{\Sigma})$.

This Gaussian measurement model simplifies real-world noise characteristics including
atmospheric correction residuals, sensor artifacts, and adjacency effects. 
This assumption is standard in remote sensing uncertainty frameworks and establishes
a statistical baseline. Importantly, the Profile likelihood-based 
framework adopted here is flexible and can readily
accommodate more complex and realistic noise structures. 
Under non-Gaussian conditions, whereas 
the expected Fisher information becomes analytically intractable,
the observed Fisher information and likelihood ratio approaches 
are still applicable within the theoretical framework presented here.

The log-likelihood can be computed as:

\begin{equation}
\ell(\boldsymbol{z}, \boldsymbol{\omega}; \textbf{y}) = -\frac{1}{2} (\textbf{y} - \boldsymbol{R}_{rs}^{S+})^T \boldsymbol{\Sigma}^{-1} (\textbf{y} - \boldsymbol{R}_{rs}^{S+}) + \text{Cte}
\end{equation}

The score for the interest parameter, \textit{i.e.}, the derivative of the log-likelihood with respect to all the parameters \(\boldsymbol{\theta}\), is expressed as follow:

\begin{equation}
\frac{\partial \ell}{\partial\boldsymbol{\theta}} = \left( \frac{\partial \boldsymbol{R}_{rs}^{S+}}{\partial \boldsymbol{\theta}} \right)^T \boldsymbol{\Sigma}^{-1} (\textbf{y} - \boldsymbol{R}_{rs}^{S+}) {\label{eq:score}}
\end{equation}

The Expected Fisher Information Matrix (EFIM) \( \boldsymbol{i}(\boldsymbol{\theta}) \) and the 
Observed Fisher Information Matrix (OFIM)  $ \boldsymbol{j}(\boldsymbol{\theta};\textbf{y})$ are given by \cite{Tocquet2001}: 

\begin{equation}
\boldsymbol{i}(\boldsymbol{\theta}) = \left( \frac{\partial \boldsymbol{R}_{rs}^{S+}}{\partial \boldsymbol{\theta}} \right)^T \boldsymbol{\Sigma}^{-1} \left( \frac{\partial \boldsymbol{R}_{rs}^{S+}}{\partial \boldsymbol{\theta}} \right)
\end{equation}

\begin{equation}
\boldsymbol{j}(\boldsymbol{\theta};\textbf{y}) = \boldsymbol{i}(\boldsymbol{\theta}) - (\textbf{y} - \boldsymbol{R}_{rs}^{S+})^T \boldsymbol{\Sigma}^{-1} \frac{\partial^2 \boldsymbol{R}_{rs}^{S+}}{\partial \boldsymbol{\theta}^2}
\end{equation}

$i_p$ and $j_p$ used for \gls{WCD} first-order inference in Wald test statistics (\ref{eq:t_e}) and (\ref{eq:t_o})
can be readily obtained as the inverse of the diagonal elements corresponding to $z$ of $\boldsymbol{i}^{-1}(\boldsymbol{\hat\theta})$ 
and $\boldsymbol{j}^{-1}(\boldsymbol{\hat\theta};\textbf{y})$. 
with respect to $z$ at the constrained \gls{MLE} $\boldsymbol{\hat\theta}_\psi$, it's straightforward to compute the 
score test (\ref{eq:s}). 


\section{Results}
We conducted a series of experiments to 
investigate the performance of \gls{WCD} inference under various environmental
setups. These experiments were carried 
out for different configurations of the parameter vector \(\boldsymbol{\tilde\theta}= \{\tilde\psi : \tilde z, \boldsymbol{\tilde\omega}  : \{\tilde{\alpha_m}, \tilde{C_{\text{chl}}},\tilde{C_{\text{cdom}}},\tilde{C_{\text{tsm}}}\}\}\),
covering different environmental conditions and sensor technologies.

\subsection{Experimental setup}

\begin{figure}[htbp]
  \centering  
  \includegraphics[width=.48\textwidth]{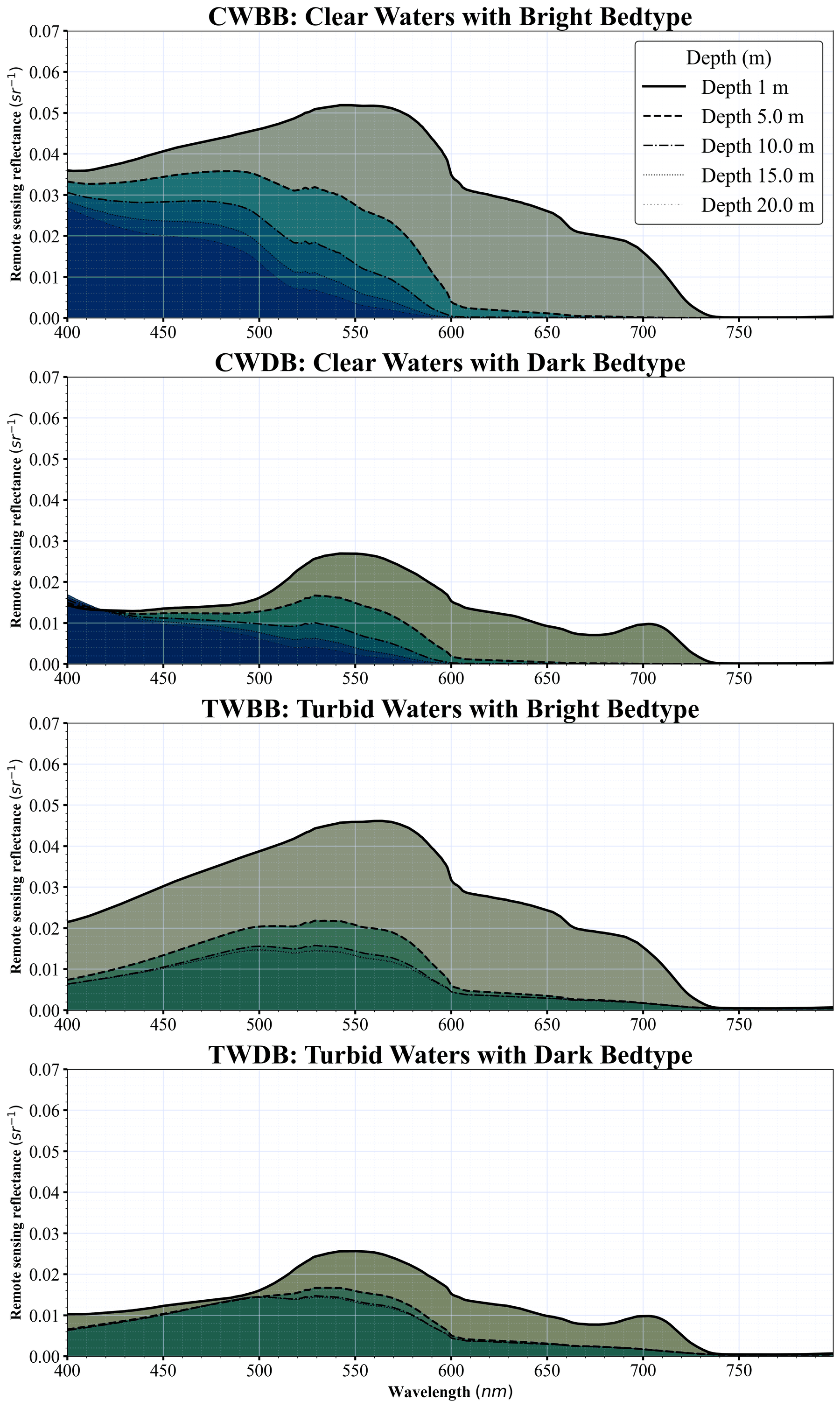}
  \caption{%
    Modeled remote sensing reflectance spectra ($R_{rs}$) 
    for water column parameterizations at varying depths (1\,m to 20\,m), 
    corresponding to Table~\ref{tab:waters}. Spectra generated using 
    bio-optical models for chlorophyll, CDOM, and TSM (Lee model). 
    Colors filling the area under each curve indicate the in-situ 
    optical appearance of the water column. 
    Colors computed using the CIE~1931 
    2$^\circ$ standard observer color matching functions \cite{ISO11664-1}, 
    converted to sRGB for display.}
  \label{fig:spectra}%
  \end{figure}

We explored distinct water conditions by adjusting the \gls{WCD} $z$ in 5-meter 
increments from 1 to 20 meters, according to different \gls{IOP} parameterizations and seabed reflectance mixing 
ratios. The experimental conditions of the water column and the seabed are
comprehensively detailed in Table \ref{tab:waters}. The parametrizations 
systematically span coastal optical diversity through a factorial design crossing 
water clarity with bottom substrate type. Clear conditions ($C_\text{chl}=0.05$\,mg/m$^3$) 
represent oligotrophic environments while turbid conditions ($C_\text{chl}=0.5$\,mg/m$^3$) represent 
mesotrophic to moderately eutrophic zones consistent with global
coastal databases~\cite{bricaud_chl,bricaud,Garcia2015,hedley,lee1999hyperspectral}.
Although higher concentrations occur in strongly eutrophic estuaries,
such conditions typically preclude reliable bottom detection and fall
outside the scope of optical bathymetric retrieval. Atmospheric variability 
was excluded intentionally to isolate the fundamental statistical inference limitations
of the water-column inversion process itself.
Parametrizations were established based on distinct aquatic and 
substrate conditions, designated as Clear Water with Bright Bedtype (\gls{CWBB}) for clear water bodies
over sandy substrates, Clear Water with Dark Bedtype (\gls{CWDB}) for clear water bodies over
vegetated or dark substrates, Turbid Water with Bright Bedtype (\gls{TWBB}) for turbid or nutrient-rich
waters over sandy bottoms, and Turbid Water with Dark Bedtype (\gls{TWDB}) for turbid waters over 
dark or organic-rich substrates. These categorizations cover a 
spectrum of chlorophyll, dissolved organic material, suspended 
solids, and albedo parameters, to represent the diverse conditions 
under which \gls{WCD} inference was evaluated.
These conditions, visually
represented in Fig. \ref{fig:spectra}, are illustrated with their spectral 
reflectances and \gls{RGB} approximations derived
from the theoretical \( R_{rs} \) values to highlight 
spectral discrepancies across parameter sets.
\begin{table}[htbp]
  \centering
  \caption[Radiometric analysis parameterizations]{True parameter values ($\tilde\theta$) of the water column used in the analysis.
   The table presents four different simulated conditions: \gls{CWBB}, \gls{TWBB}, \gls{CWDB}, and \gls{TWDB}.}\label{tab:waters}

  \begin{tblr}{
    width = \linewidth,
    colspec = {Q[115]Q[210]Q[177]Q[192]Q[223]},
    cells = {c},
    row{3} = {melrose},
    row{5} = {melrose},
    cell{1}{2} = {melrose},
    cell{1}{3} = {melrose},
    cell{1}{4} = {melrose},
    cell{1}{5} = {melrose},
    vline{2-6} = {1}{melrose},
    hline{1} = {2-5}{0.08em},
    hline{2} = {-}{0.08em},
    hline{6} = {-}{},
  }
  & $C_{\text{chl}}$ ($mg/m^3$) & $C_{\text{cdom}}$ ($m^{-1}$) & $C_{\text{tsm}}$ ($g/m^3$) & $\alpha_M$ \\
  \textbf{\gls{CWBB}} & 0.05               & 0.05              & 0.1               & 0.8        \\
  \textbf{\gls{TWBB}} & 0.5                & 0.5               & 3                 & 0.8        \\
  \textbf{\gls{CWDB}} & 0.05               & 0.05              & 0.1               & 0.1        \\
  \textbf{\gls{TWDB}} & 0.5                & 0.5               & 3                 & 0.1        
  \end{tblr}
\end{table}

Monte Carlo simulations and optimization
procedures were conducted as outlined in Table \ref{tab:exp_config_column}. 
The experiments spanned a wavelength range of 
400-800 nm with a spectral resolution of 1 nm, adhering to the 
spectral range commonly used in coastal optical remote sensing.
These procedures employed 1000 iterations for Monte Carlo simulations and
leveraged the trust-region method from the SciPy library for optimization \cite{trf}, 
selected for its optimal performance in our testing scenarios.
For each sample, profiles were computed using a regularly 
spaced grid around the \gls{MLE} comprised of 30 profiling points.
The variance for modeling the reflectance noise was set to 
\(1 \cdot 10^{-7}\), which is a representative value of high-end
satellite imagers in the coastal cover \cite{jay}.
\begin{table}[htbp]
\centering
\caption[Operational configuration for the radiometric \gls{WCD} inference under \gls{SA} model]{Experimental configuration.}
\label{tab:exp_config_column}
\begin{tblr}{
  width = \linewidth,
  colspec = {Q[438]Q[465]},
  cells = {c},
  row{odd} = {melrose},
  hline{1-2,8} = {-}{},
}
\textbf{Parameter}  & \textbf{Value}   \\
Noise (R\(_{rs}\) variance)    & $1 \cdot 10^{-7}$ \\
Wavelength Range    & 400-800nm        \\ 
Spectral resolution & 1nm              \\
Optimizer           & Trust-region     \\
Monte Carlo Samples & 1000           \\  
Profiling points   & 30          
\end{tblr}
\end{table}
The optimization approach without 
boundary constraints was chosen to estimate parameters
of the genuine likelihood function that corresponds
to the actual data-generating process, 
rather than a modified
likelihood that implicitly incorporates 
uniform prior assumptions about parameter ranges.
Initiating the experiments from true 
parameter values facilitated swift
convergence to a viable local log-likelihood maximum. 
To highlight the significance 
of spectral data granularity on \gls{WCD} inference 
accuracy, our study further differentiated between 
two sensor technologies.
We analyzed multispectral
data captured by the Sentinel \gls{MSI-2A} sensor (7 bands across the spectral domain 400-800nm)
alongside hyperspectral data obtained from the
hySpex sensor enabling a deeper understanding of how sensor
resolution influences \gls{WCD} inference outcomes.

\subsection{\gls{WCD} profiling results}

Having established the experimental configurations 
and the analytical framework, we now proceed 
to examine the profiling process and its impact on \gls{WCD} inference
accuracy across varying environmental conditions.

\begin{figure*}[!htbp]
  \centering
\subfloat[]{\includegraphics[width=0.45\textwidth]{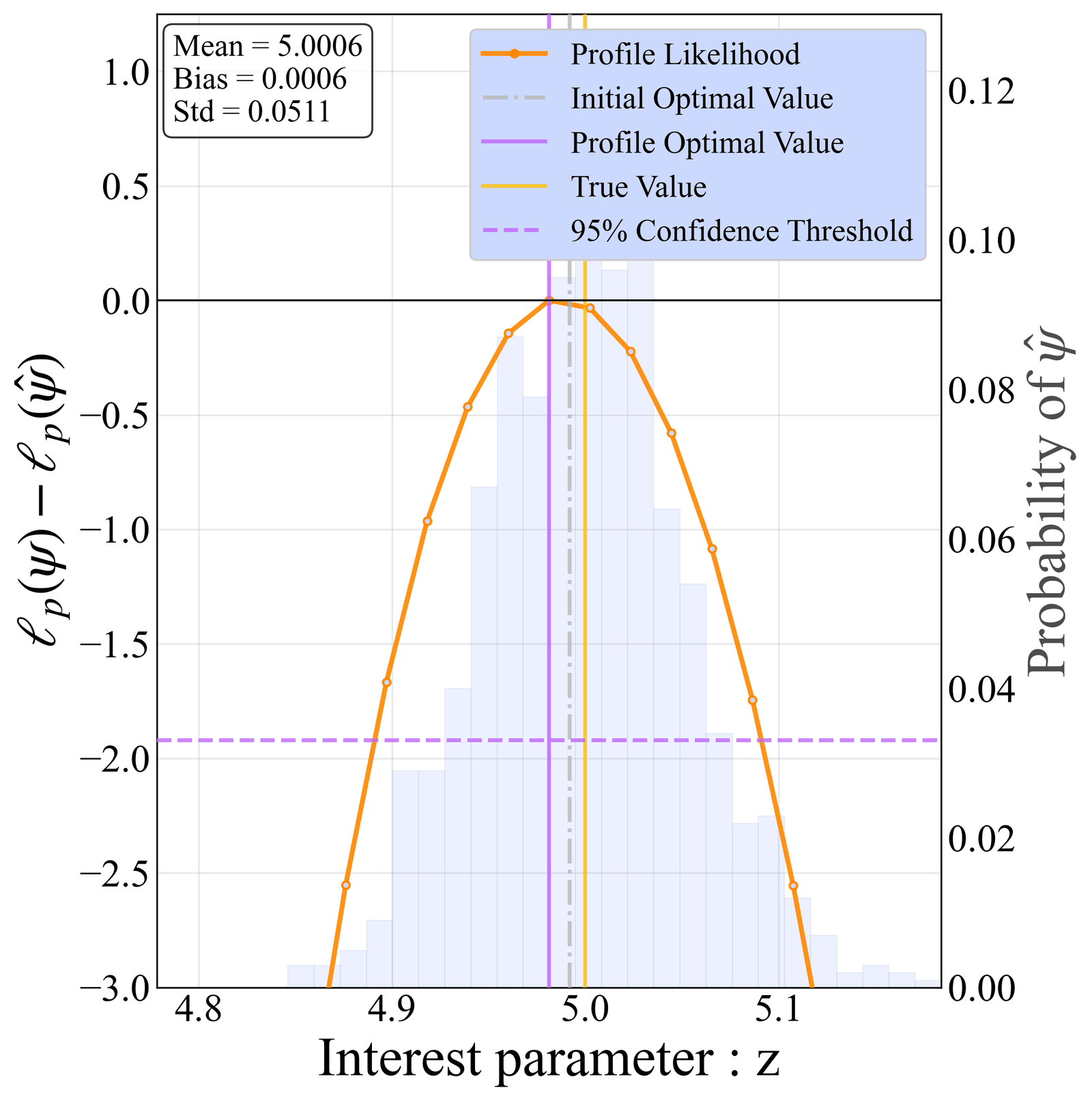}%
\label{fig:ideal_case}}
\hfil
\subfloat[]{\includegraphics[width=0.45\textwidth]{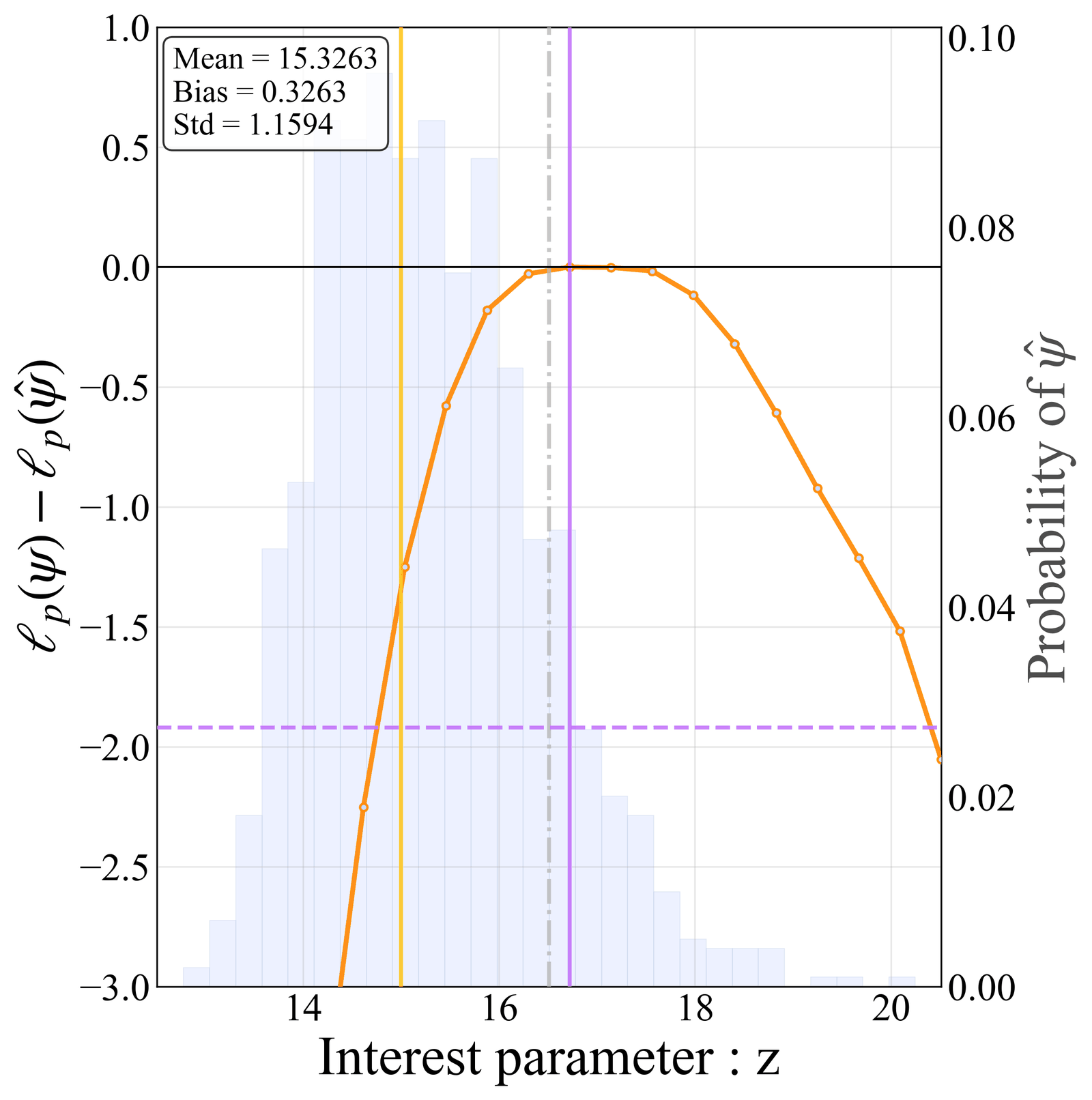}%
\label{fig:asymmetric_case}}
\hfil
\subfloat[]{\includegraphics[width=0.45\textwidth]{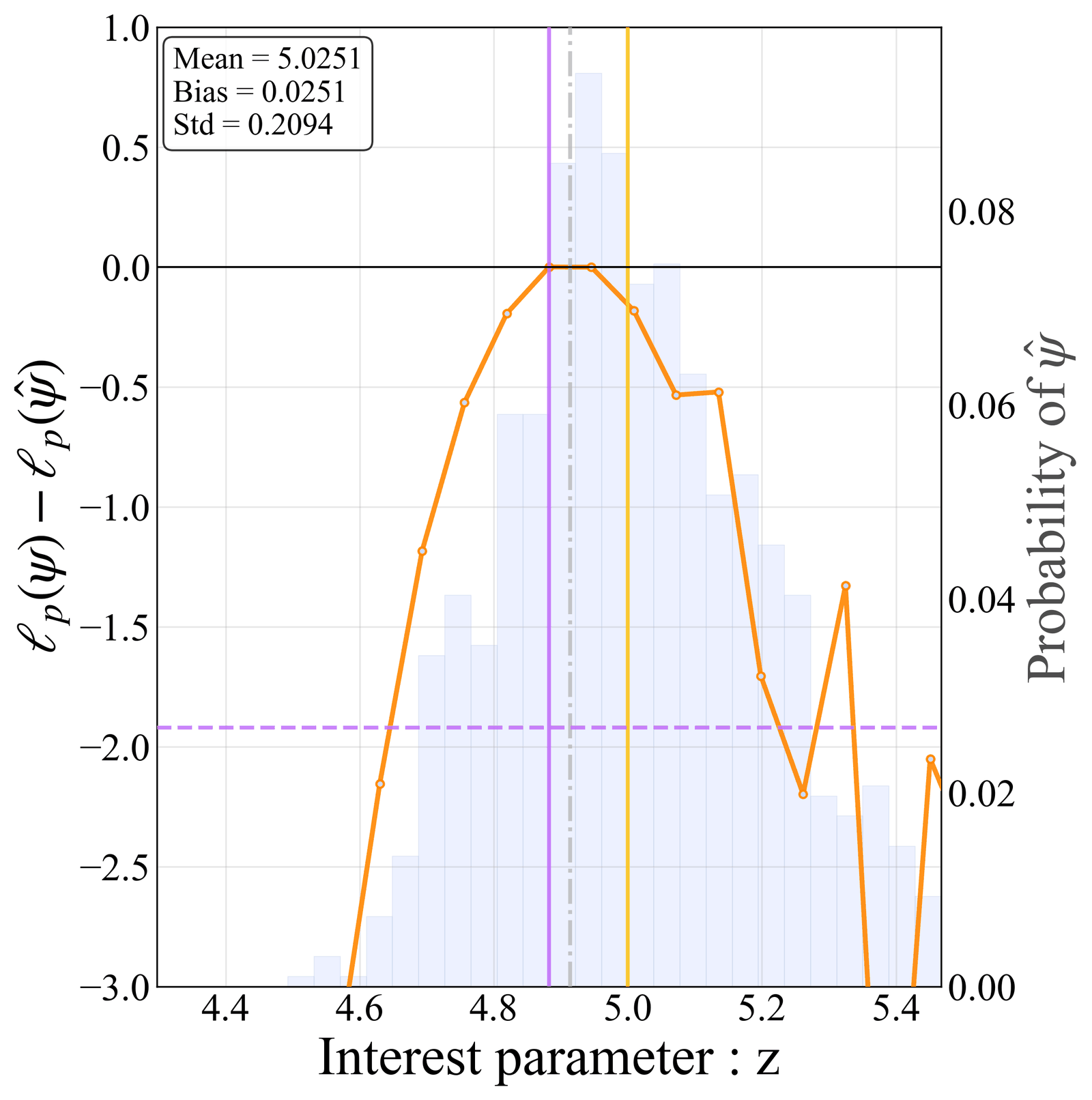}%
\label{fig:outlier_case}}
\hfil 
\subfloat[]{\includegraphics[width=0.45\textwidth]{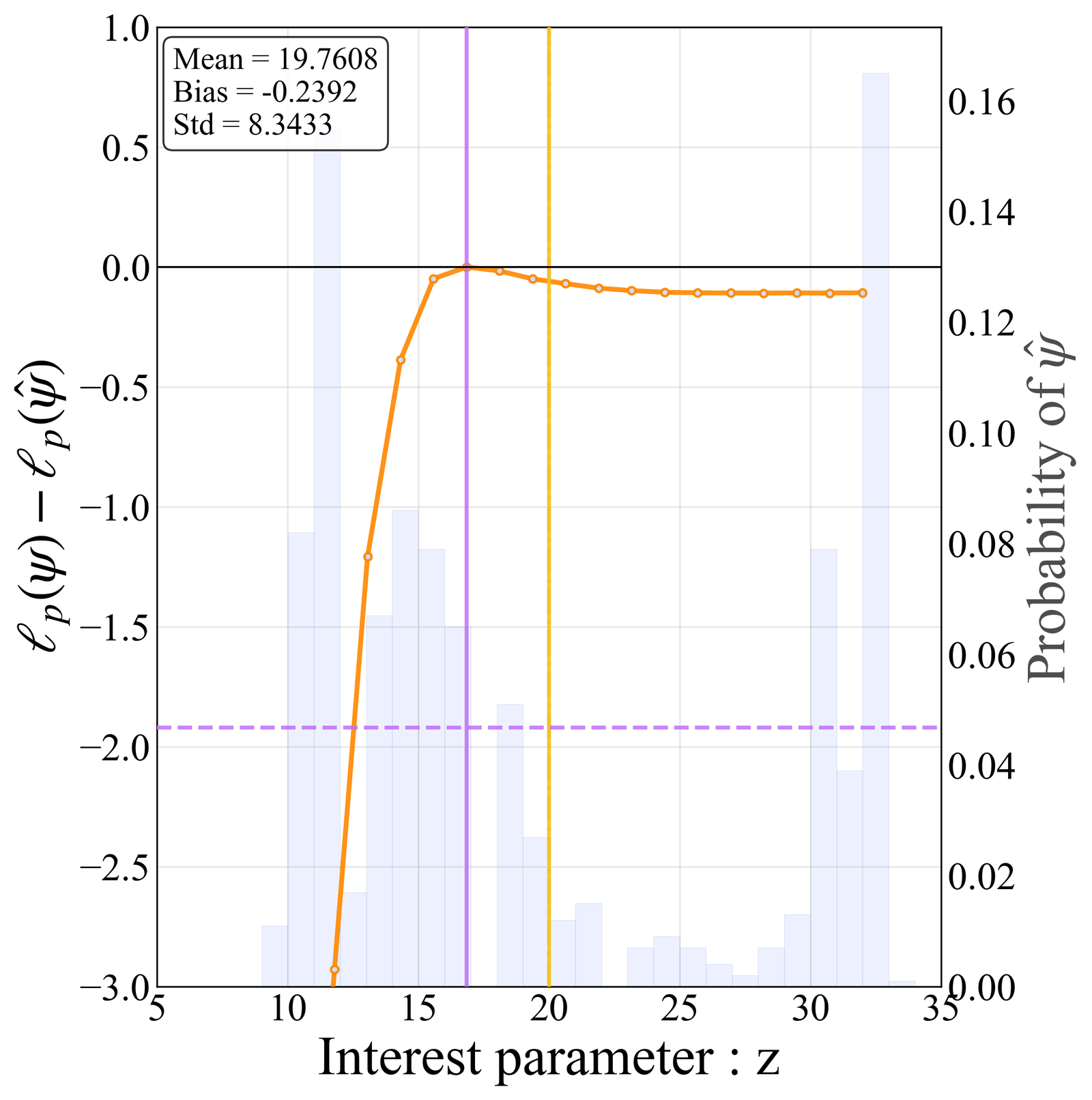}%
\label{fig:practical_identifiability}}
\caption{Profile likelihood curves (orange color) for different case studies.
 (a) Ideal case: \gls{CWBB} at 5m (hyperspectral). (b) Asymmetric case: \gls{CWDB} at 15m
  (multispectral). (c) Outlier case: \gls{CWBB} at 5m (multispectral). (d) Practical identifiability: \gls{TWDB} at 20m (hyperspectral). 
  Vertical lines represent the true \gls{WCD} value (yellow), 
  as well as the initial (grey) and refined (melrose) \gls{WCD} estimates. 
  The statistical distribution of the \gls{WCD} \gls{MLE} is
   overlayed (light blue) in the right axis with descriptive statistics (Mean, Bias, Standard deviation (Std)).}
\label{fig:case_studies}
\end{figure*}

Our profiling analysis revealed several distinct behavioral patterns that 
illustrate the performance of likelihood-based inference in \gls{SA} models 
for \gls{WCD} estimation. These patterns emerge systematically across different
environmental conditions and sensor configurations, with their manifestation 
depending on the interplay between sensor spectral resolution, water optical 
complexity, and depth.

\begin{itemize}
\item{\textbf{Ideal Behavior Pattern:}}
Under favorable conditions—typically combining high spectral resolution
with optically simple waters at moderate depths—we observe symmetric 
parabolic likelihood curves around the \gls{MLE}, indicating well-behaved 
likelihood surfaces (exemplified in Fig.~\ref{fig:ideal_case} for hyperspectral
CWBB at 5m). This symmetric profiles pattern demonstrates
the validity of large-sample theory in providing robust uncertainty quantification. 
We observed that it predominates in hyperspectral data for clear waters 
up to 10-15m depth, and in multispectral data only for very shallow clear waters
(1-5m).

\item{\textbf{Asymmetric Behavior Pattern:}}
As environmental complexity increases or sensor resolution decreases, 
asymmetry emerges in the likelihood surfaces. This pattern manifests 
systematically across scenarios: in hyperspectral data under unfavorable 
optical conditions (\gls{TWBB} and \gls{TWDB}) at depths greater than 10m,
and more extensively in multispectral data where significant asymmetry 
appears at shallower depths (5m) even in clear waters 
(illustrated in Fig.~\ref{fig:asymmetric_case}). 
This asymmetry fundamentally influences \gls{CI} shapes as shown further 
in the following section \ref{sec:ci_characterization} (Fig.\ref{fig:violin_patterns}), 
with $r$ and $s$ tests producing asymmetric 95\% \gls{CI}s that 
more accurately reflect the \gls{MLE} distribution, while Wald tests
($t_e$, $t_o$) produce potentially misleading symmetric intervals.

\item{\textbf{Outlier Pattern:}}
Irregular likelihood surfaces with outliers disrupting expected 
parabolic characteristics represent a systematic challenge that 
intensifies with reduced spectral information and increased environmental 
complexity. While observable across various scenarios, this pattern
becomes particularly pronounced in multispectral data, 
even in relatively benign conditions such as shallow clear waters
(demonstrated in Fig.~\ref{fig:outlier_case}).
These irregular profile likelihood shapes resulted in multiple 
intersection points on statistical
test curves for the computation of confidence limits. 

\item{\textbf{Practical Identifiability Limitations:}}
Extreme asymmetry leading to practical identifiability issues 
represents the most challenging scenario for first-order inference. 
This pattern emerges systematically when the combination of limited
spectral information, optical complexity, and \gls{WCD} approaches the 
physical limits of parameter estimation. While most severely demonstrated 
in multispectral data for challenging scenarios like \gls{TWDB} at greater
depths, the onset varies predictably: for hyperspectral data, the upper bound identifiability
issues primarily arise in the most extreme conditions (15m-20m,\gls{TWDB}-\gls{TWBB})
as exemplified in Fig.~\ref{fig:practical_identifiability},
whereas for multispectral data, these limitations 
can emerge as early as 10m depth across various parametrizations. 
Non-identifiable upper bounds for both likelihood ratio and
score tests emphasize the fundamental challenges of interpreting 
first-order 95\% \gls{CI}s when approaching the small-sample regime limits.
This pattern manifests when Fisher information becomes critically low, often 
characterized by near-singular Fisher information matrices or condition numbers
exceeding practical thresholds, indicating insufficient spectral information 
for reliable parameter estimation.

\end{itemize}

Notably, the profiling process demonstrated robustness in handling
local optimization challenges inherent in SA models. 
As illustrated in Fig. \ref{fig:case_studies}, the profiling procedure -
which re-optimizes nuisance parameters for each fixed \gls{WCD} value -
effectively corrected suboptimal initial \gls{MLE} estimates (shown by the transition 
from initial estimates in dashed gray to refined estimates in melrose color). 
In what follows, we consider these refined estimates as the effective \gls{MLE} values,
 since they better represent the maximization outcome than the initial estimates.
Even under our idealized experimental conditions with true parameter 
initialization, this iterative refinement enhanced both parameter 
estimation accuracy and uncertainty quantification reliability, 
particularly for Wald test-based \gls{CI}.

These behavioral patterns form a systematic hierarchy of inference
reliability, with sensor spectral resolution and environmental 
optical complexity serving as primary determinants of which pattern
dominates in any given scenario. 

To mitigate the risk of underestimating uncertainty in cases with multiple 
intersections, we consistently chose the farthest intersection point in our analyses. 
For cases where no intersection was found, such as in practical identifiability scenarios,
we considered the domain limits (0-35m) and hence assigned no rejection, 
which represents the safest approach for uncertainty quantification.

These case studies highlight the complexities of \gls{WCD} inference 
across different environmental conditions and sensor types, underscoring 
the importance of careful interpretation of uncertainty estimates in
\gls{SA} models.

\subsection{\gls{WCD} inference results}

This section evaluates the performance of first-order statistical tests
($r$, $t_o$, $t_e$, and $s$) in estimating 95\% \gls{CI} for \gls{WCD} and their coverage probabilities.

\subsubsection{Confidence Intervals Characterization\label{sec:ci_characterization}}

\begin{figure*}[!t]
\centering
\subfloat[Ideal Case: 5m CWBB hySpex]{\includegraphics[width=0.48\textwidth]{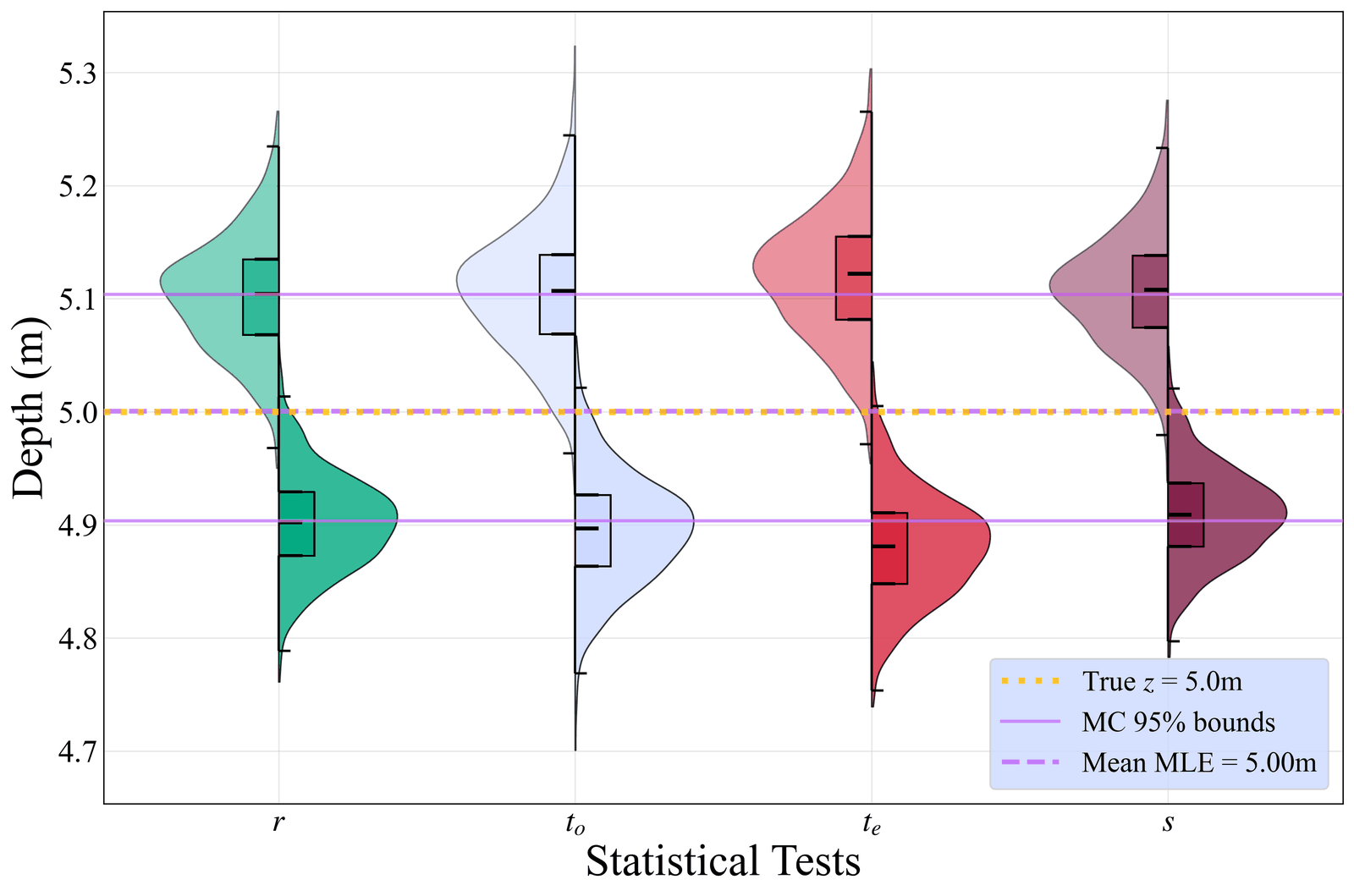}%
\label{fig:violin_ideal}}
\hfil
\subfloat[Asymmetric Case: 15m CWDB Sentinel]{\includegraphics[width=0.48\textwidth]{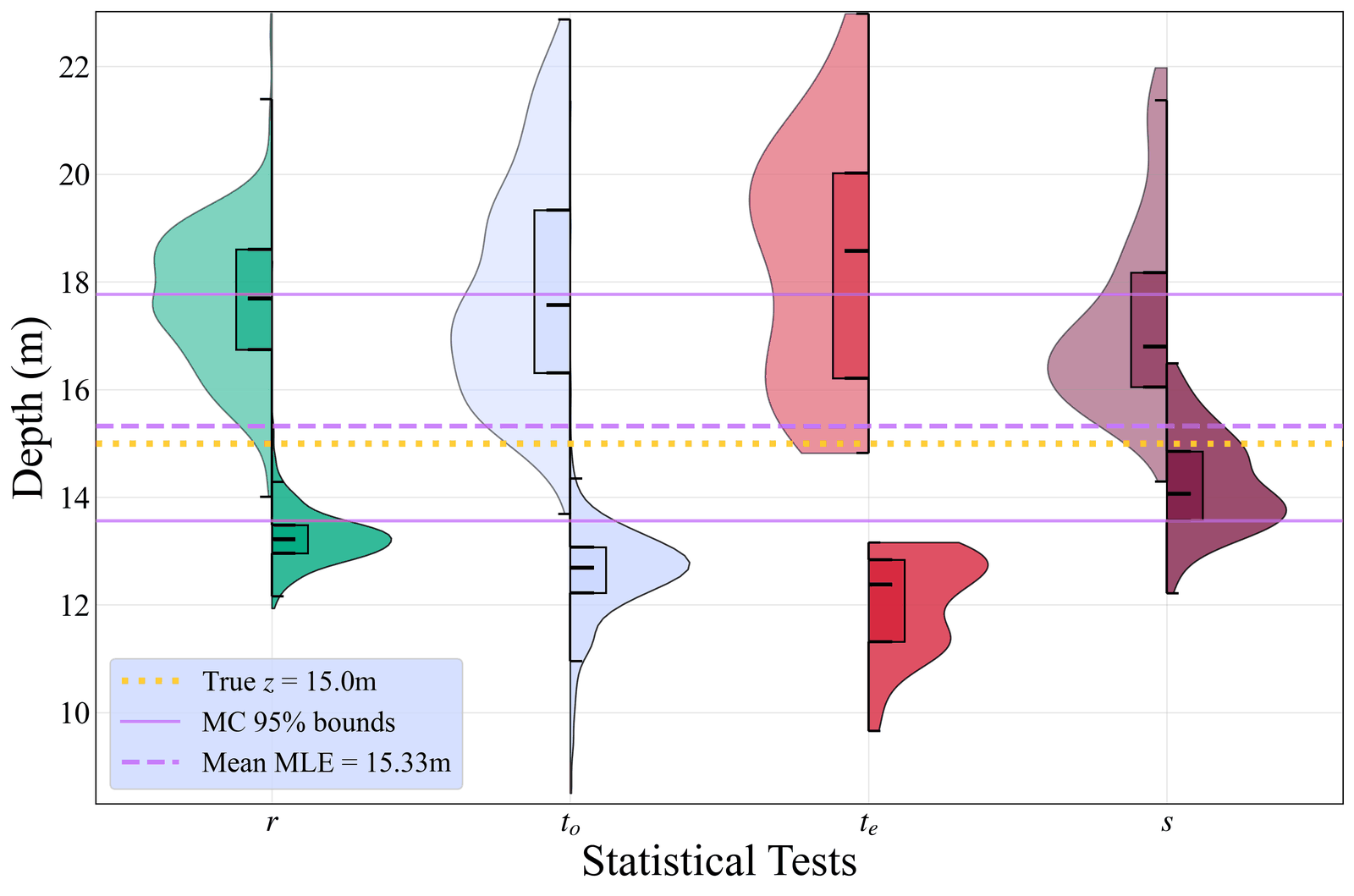}%
\label{fig:violin_asymmetric}}
\hfil
\subfloat[Outlier Case: 5m CWBB Sentinel]{\includegraphics[width=0.48\textwidth]{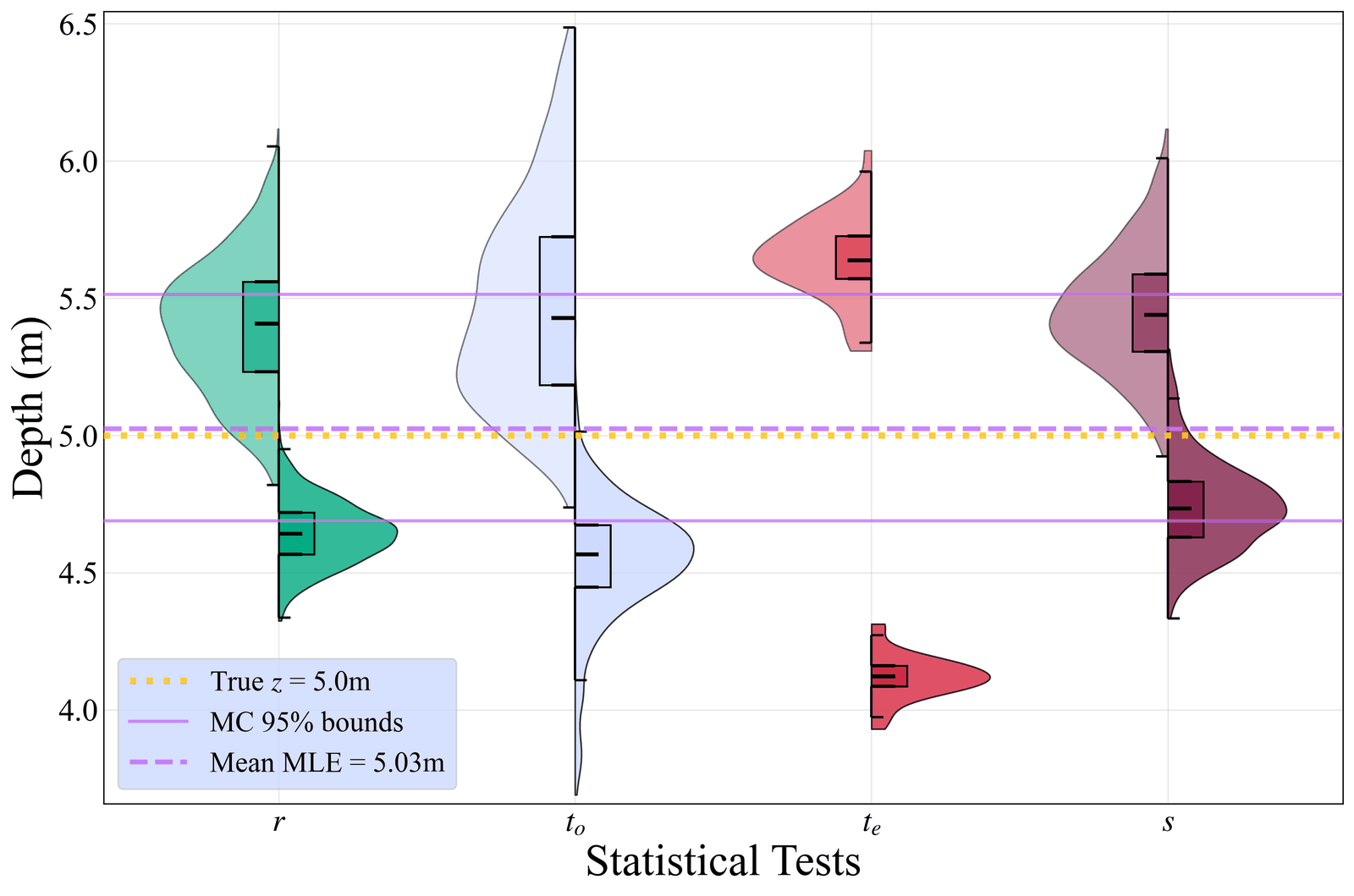}%
\label{fig:violin_outlier}}
\hfil 
\subfloat[Identifiability Case: 20m TWDB hySpex]{\includegraphics[width=0.48\textwidth]{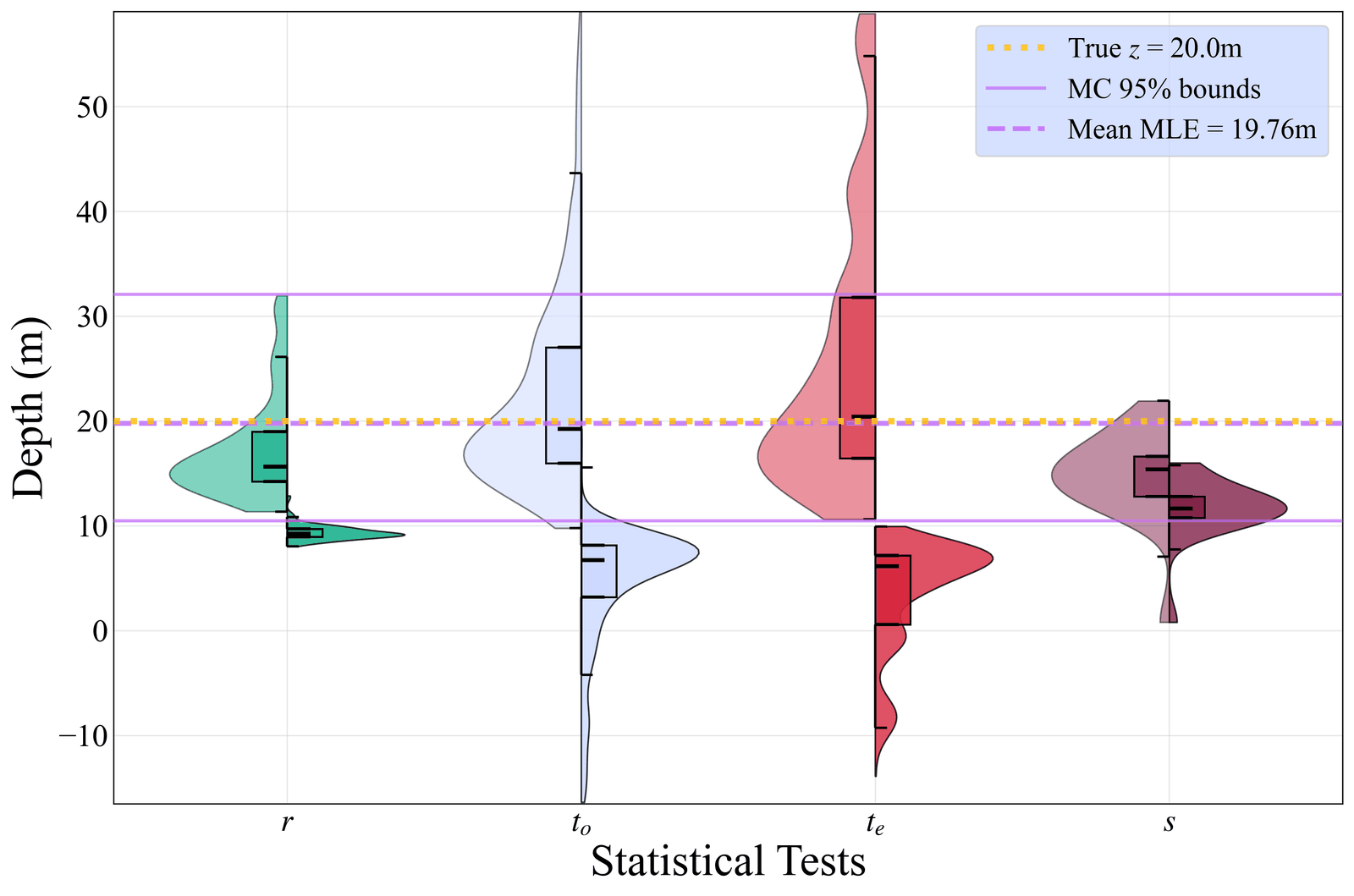}%
\label{fig:violin_identifiability}}
\caption{Confidence intervals distribution patterns across representative scenarios. 
Each violin combines upper and lower bounds distributions, with reference lines showing 
true \gls{WCD} (yellow dotted). The Melrose
horizontal lines denote the Monte Carlo (MC) derived 
average \gls{MLE} (dashed) and the corresponding
95\% quantiles (solid).
The patterns demonstrate increasing statistical challenges from ideal hyperspectral conditions 
to problematic multispectral configurations.}
\label{fig:violin_patterns}
\end{figure*}

Before presenting quantitative rejection rate analysis, we examine the underlying 
confidence interval distributions to understand the statistical behavior across 
different scenarios. Figure \ref{fig:violin_patterns} shows representative cases 
for each behavioral pattern identified in the profiling analysis, revealing the 
distribution of confidence interval bounds for all statistical tests.

These distributions reveal several critical insights into the performance
of statistical tests across different environmental and sensor 
configurations. In the hyperspectral context for the \gls{CWBB} 
case at 5m  (Fig.~\ref{fig:violin_ideal}), 
all statistical tests produce well-centered, symmetric distributions 
around the true \gls{WCD} value, with the expected Fisher information 
Wald test ($t_e$) showing slightly wider
 95\% \gls{CI} compared to the empirical quantiles.
The multispectral scenario for the \gls{CWDB} case 
at 15m (Fig.~\ref{fig:violin_asymmetric}) demonstrates
how environmental complexity begins to challenge the symmetry
assumptions of Wald tests, while likelihood ratio ($r$) 
and score ($s$) tests adapt better to the non-linear
likelihood surface.

More critically, the analysis reveals the systematic degradation
of the expected Fisher information approach ($t_e$, equivalent 
to variance-covariance propagation) across challenging scenarios.
In the multispectral scenario for the \gls{CWBB} case at 5m (Fig.~\ref{fig:violin_outlier}), $t_e$ 
produces excessively liberal \gls{CI}, visible
as artificially wide distributions that substantially 
overestimate uncertainty. This pattern becomes particularly 
pronounced in multispectral applications, where the method 
fails rapidly beyond 5m water clumn depth even in clear waters.

The hyperspectral scenario for the \gls{TWDB} case at 20m (Fig.~\ref{fig:violin_identifiability}) 
exemplifies the most challenging scenarios where optimization
hurdles and parameter identifiability issues lead to degenerate
\gls{CI} for some tests. Such cases, while 
informative about fundamental model limitations, cannot
provide reliable rejection rate statistics for comparative 
analysis and are excluded from the quantitative assessment of coverage probabilities
that follows.

Notably, across all scenarios, the likelihood ratio test ($r$)
demonstrates superior robustness, maintaining reasonable 
interval distributions even under challenging conditions,
while the observed Fisher information Wald test ($t_o$) 
generally outperforms its expected counterpart ($t_e$) 
by adapting to local likelihood curvature rather than 
relying on theoretical average information.

\subsubsection{Rejection Rate Analysis}

For robust statistical comparison, we focus on scenarios
where \gls{CI}
are well-behaved and physically meaningful. 
95\% \gls{CI} should ideally reject 
the true value 5\% across  n=1000 Monte Carlo samples.
We categorize the rejection rates  based on 
statistical theory for the Monte Carlo samples as follows:
\begin{itemize}
    \item \textbf{4.0\% - 6.0\%}: Optimal (well-calibrated CI)
    \item \textbf{3.0\% - 7.0\%}: Acceptable (within expected statistical variation)
    \item \textbf{$<$ 3.0\%}: Liberal (\gls{CI} wider than ideal, uncertainty overestimation)
    \item \textbf{$>$ 7.0\%}: Conservative (\gls{CI} narrower than expected, uncertainty underestimation)
\end{itemize}

Tables \ref{tab:rej_rate_hys} and 
\ref{tab:rej_rate_sen} present rejection rates for hyperspectral 
and multispectral sensors respectively, with cases marked as N/A where 
\gls{CI} were degenerate due to optimization challenges and outliers, 
preventing reliable rejection rate computation.


Several critical findings emerge from these tables analysis. Hyperspectral
data consistently demonstrates superior performance for first-order 
statistical inference, with rejection rates generally aligning with 
the nominal 5\% value across most scenarios. The likelihood ratio test ($r$)
proves most robust across all conditions, while the expected
Fisher information Wald test ($t_e$) confirms the effectiveness of
variance-covariance propagation approaches in hyperspectral settings 
under favorable conditions.

In contrast, multispectral data presents substantial challenges, particularly
in complex environmental conditions. Performance degradation occurs at
shallower depths (5m) even in clear waters, with significant deviations
from nominal rejection rates in turbid environments. 
The expected Fisher information Wald test ($t_e$) is generally outperformed
by the observed information variant ($t_o$) and the score test ($s$) in most scenarios,
while all tests struggle compared to the likelihood ratio approach.

The systematic failure of the expected Fisher information approach 
($t_e$) in multispectral applications is particularly noteworthy, 
as this method is equivalent to classical variance-covariance uncertainty 
propagation widely used in remote sensing. 
A liberal behavior for $t_e$-based \gls{CI} is observed 
in confidence interval distributions, especially in clear waters,
(excessive widths leading to low
rejection rates). This failure stems from a fundamental statistical constraint:
the multispectral sensor with 7 bands must 
simultaneously estimate 5 unknown parameters
(depth, water constituents, bottom properties), leaving insufficient statistical 
degrees of freedom for robust uncertainty quantification. The resulting parameter
correlations and near-singular Fisher information matrices drive traditional 
variance-covariance approaches to produce unrealistically wide \gls{CI}.

These results have important practical implications. While hyperspectral 
sensors enable reliable uncertainty quantification across diverse environmental
conditions, multispectral applications require careful interpretation,
particularly beyond shallow, clear waters. The findings underscore
the superior statistical foundation provided by hyperspectral measurements
for operational bathymetric uncertainty assessment and highlight 
the need for alternative uncertainty quantification strategies 
in multispectral applications.

\definecolor{optimal}{RGB}{0,128,0}    
\definecolor{liberal}{RGB}{255,140,0}   
\definecolor{conservative}{RGB}{220,20,60} 

\newcommand{\colorrate}[1]{%
  \ifdim#1pt<3pt%
    \textcolor{liberal}{#1}%
  \else%
    \ifdim#1pt>7pt%
      \textcolor{conservative}{#1}%
    \else%
      \ifdim#1pt>4pt%
        \ifdim#1pt<6pt%
          \textcolor{optimal}{\textbf{#1}}%
        \else%
          #1%
        \fi%
      \else%
        #1%
      \fi%
    \fi%
  \fi%
}

\begin{table}[!t]
\centering
\caption[First-order statistical tests rejection rates for the hySpex sensor.]{First-order
 statistical tests performance for the hySpex sensor. 
Rejection rates are color-coded: \textcolor{optimal}{green} indicates optimal performance, 
\textcolor{liberal}{orange} indicates liberal response (uncertainty overestimation), 
and \textcolor{conservative}{red} indicates conservative response (uncertainty underestimation).
N/A indicates cases where confidence bounds were unreliable.}
\label{tab:rej_rate_hys}
\begin{tblr}{
width = \linewidth,
colspec = {Q[158]Q[263]Q[110]Q[144]Q[144]Q[110]},
row{even} = {melrose,c},
row{1} = {melrose},
row{3} = {c},
row{5} = {c},
row{7} = {c},
row{9} = {c},
row{11} = {c},
row{13} = {c},
row{15} = {c},
row{17} = {c},
row{19} = {c},
row{21} = {c},
cell{1}{1} = {r=2}{},
cell{1}{2} = {r=2}{},
cell{1}{3} = {c=4}{0.508\linewidth,c},
cell{3}{1} = {r=4}{},
cell{7}{1} = {r=4}{},
cell{11}{1} = {r=4}{},
cell{15}{1} = {r=4}{},
cell{19}{1} = {r=4}{},
hline{1,3,23} = {-}{0.08em},
hline{2} = {3-6}{},
hline{7,11,15,19} = {-}{0.05em},
}
\textbf{Depth (m)} & \textbf{Param.} & \textbf{Rejection rates (\%)} &       &              &               \\
                &                           & $r$                           & $t_o$ & $t_e$        & $s$           \\
\textbf{1.0}    & \gls{CWBB}                      & \colorrate{4.9}                           & \colorrate{5.8}   & \colorrate{0.8}         & \colorrate{5.7}           \\
                & \gls{CWDB}                      & \colorrate{5.1}                           & \colorrate{6.8}   & \colorrate{2.0}         & \colorrate{4.9}           \\
                & \gls{TWBB}                      & \colorrate{5.3}                           & \colorrate{5.3}   & \colorrate{5.4}         & \colorrate{4.9}         \\
                & \gls{TWDB}                      & \colorrate{6.7}                           & \colorrate{4.9}   & \colorrate{4.9}         & \colorrate{5.0}  \\
\textbf{5.0}    & \gls{CWBB}                      & \colorrate{3.1}                           & \colorrate{4.8}   & \colorrate{2.4}         & \colorrate{3.0}          \\
                & \gls{CWDB}                      & \colorrate{4.1}                           & \colorrate{4.7}   & \colorrate{2.5}         & \colorrate{3.5}          \\
                & \gls{TWBB}                      & \colorrate{5.7}                           & \colorrate{6.5}   & \colorrate{4.9}         & \colorrate{4.4}          \\
                & \gls{TWDB}                      & \colorrate{4.6}                           & \colorrate{5.2}   & \colorrate{3.3}         & \colorrate{4.0}           \\
\textbf{10.0}   & \gls{CWBB}                      & \colorrate{5.2}                           & \colorrate{6.3}   & \colorrate{4.5}         & \colorrate{4.2}           \\
                & \gls{CWDB}                      & \colorrate{6.3}                           & \colorrate{5.5}   & \colorrate{3.9}         & \colorrate{4.7}           \\
                & \gls{TWBB}                      & \colorrate{10.3}                          & \colorrate{12.4}  & \colorrate{9.2}         & \colorrate{4.9}           \\
                & \gls{TWDB}                      & \colorrate{7.0}                           & \colorrate{13.7}  & \colorrate{9.7}         & \colorrate{7.9}           \\
\textbf{15.0}   & \gls{CWBB}                      & \colorrate{5.3}                           & \colorrate{5.6}   & \colorrate{3.1}         & \colorrate{4.8} \\
                & \gls{CWDB}                      & \colorrate{4.0}                           & \colorrate{5.1}   & \colorrate{2.0}         & \colorrate{4.7}           \\
                & \gls{TWBB}                      & N/A    & N/A & N/A & N/A \\
                & \gls{TWDB}                      & \colorrate{3.9}                           & \colorrate{16.6}  & \colorrate{15.9}        & \colorrate{4.7}         \\
\textbf{20.0}   & \gls{CWBB}                      & \colorrate{5.0}                           & \colorrate{6.3}   & \colorrate{2.3}         & \colorrate{5.0}          \\
                & \gls{CWDB}                      & \colorrate{5.9}                           & \colorrate{7.9}   & \colorrate{3.5}         & \colorrate{5.7}           \\
                & \gls{TWBB}                      & \colorrate{8.6}                           & N/A & N/A & N/A \\
                & \gls{TWDB}                      & N/A    & N/A & N/A & N/A \\
\end{tblr}
\end{table}

\begin{table}[!t]
\centering
\caption{First-order statistical tests rejection rates for the \gls{MSI-2A} sensor. 
Rejection rates are color-coded: \textcolor{optimal}{green} indicates optimal performance, 
\textcolor{liberal}{orange} indicates liberal response (uncertainty overestimation), 
and \textcolor{conservative}{red} indicates conservative response (uncertainty underestimation).
N/A indicates cases where confidence bounds were unreliable.}
\label{tab:rej_rate_sen}
\begin{tblr}{
width = \linewidth,
colspec = {Q[158]Q[263]Q[110]Q[144]Q[144]Q[110]},
row{even} = {Melrose,c},
row{1} = {Melrose},
row{3} = {c},
row{5} = {c},
row{7} = {c},
row{9} = {c},
row{11} = {c},
row{13} = {c},
row{15} = {c},
row{17} = {c},
row{19} = {c},
row{21} = {c},
cell{1}{1} = {r=2}{},
cell{1}{2} = {r=2}{},
cell{1}{3} = {c=4}{0.508\linewidth,c},
cell{3}{1} = {r=4}{},
cell{7}{1} = {r=4}{},
cell{11}{1} = {r=4}{},
cell{15}{1} = {r=4}{},
cell{19}{1} = {r=4}{},
hline{1} = {-}{},
hline{2} = {3-6}{},
hline{3,23} = {-}{0.08em},
hline{7,11,15,19} = {-}{0.05em},
}
\textbf{Depth (m)} & \textbf{Param.} & \textbf{Rejection Rates (\%)} &               &               &               \\
              &                           & $r$                           & $t_o$         & $t_e$         & $s$           \\
\textbf{1.0}  & \gls{CWBB}                      & \colorrate{4.2}                           & \colorrate{4.5}          & \colorrate{1.2}            & \colorrate{6.0}           \\
              & \gls{CWDB}                      & \colorrate{6.1}                           & \colorrate{5.3}           & \colorrate{2.3}           & \colorrate{7.3}         \\
              & \gls{TWBB}                      & \colorrate{5.1}                           & \colorrate{4.5}           & \colorrate{4.2}           & \colorrate{5.1}           \\
              & \gls{TWDB}                      & \colorrate{6.8}                           & \colorrate{4.8}          & \colorrate{4.3}          & \colorrate{5.2}           \\
\textbf{5.0}  & \gls{CWBB}                      & \colorrate{3.7}                           & \colorrate{8.4}         & \colorrate{0.0}  & \colorrate{7.6}           \\
              & \gls{CWDB}                      & \colorrate{4.3}                           & \colorrate{9.0}          & \colorrate{0.0}  & \colorrate{8.0}         \\
              & \gls{TWBB}                      & N/A    & N/A & N/A & N/A \\
              & \gls{TWDB}                      & \colorrate{6.8}                  & \colorrate{11.3}  & \colorrate{1.3}  & \colorrate{5.7} \\
\textbf{10.0} & \gls{CWBB}                      & \colorrate{4.1}                 & \colorrate{6.1}          & \colorrate{0.0} & \colorrate{9.6} \\
              & \gls{CWDB}                      & \colorrate{4.9}                  & \colorrate{4.9}          & \colorrate{0.8} & \colorrate{11.4} \\
              & \gls{TWBB}                      & N/A    & N/A & N/A & N/A \\
              & \gls{TWDB}                      & \colorrate{4.9}                & N/A & N/A & N/A \\
\textbf{15.0} & \gls{CWBB}                      & \colorrate{3.6}             & \colorrate{4.9}        & \colorrate{0.1} & \colorrate{10.6} \\
              & \gls{CWDB}                      & \colorrate{1.8}                 & \colorrate{4.6}           & \colorrate{0.1} & \colorrate{12.9} \\
              & \gls{TWBB}                      & N/A    & N/A & N/A & N/A \\
              & \gls{TWDB}                      & N/A    & N/A & N/A & N/A \\
\textbf{20.0} & \gls{CWBB}                      & \colorrate{2.1}                  & \colorrate{7.6}          & \colorrate{0.0} & \colorrate{12.9} \\
              & \gls{CWDB}                      & \colorrate{3.8}                & \colorrate{9.7}     & \colorrate{0.2} & \colorrate{15.9}\\
              & \gls{TWBB}                      & N/A    & N/A & N/A & N/A \\
              & \gls{TWDB}                      & N/A    & N/A & N/A & N/A \\
\end{tblr}
\end{table}

Several critical findings emerge from this analysis. Hyperspectral data 
consistently demonstrates superior performance for first-order statistical 
inference, with rejection rates generally aligning with the nominal 5\% value 
across most scenarios. The likelihood ratio test ($r$) proves most robust across
all conditions, while the expected Fisher information Wald test ($t_e$) 
confirms the effectiveness of variance-covariance propagation approaches 
in hyperspectral settings.

In contrast, multispectral data presents substantial challenges, particularly
in complex environmental conditions. Performance degradation occurs at shallower
depths (5m) even in clear waters, with significant deviations from nominal
rejection rates in turbid environments. The observed Fisher information
Wald test ($t_o$) generally outperforms the expected information variant
($t_e$) in challenging multispectral scenarios, though they both 
struggle compared to the likelihood ratio approach.

\section{Discussion}

The findings in this study provide a detailed exploration of likelihood-based
inference for evaluating bathymetric uncertainties within \gls{SA} models,
with emphasis on environmental factors and sensor spectral resolution effects
on \gls{WCD} uncertainties.

Earlier work by \cite{jay} indicated that for sensors like Sentinel \gls{MSI-2A},
variance-covariance bounds can dramatically increase beyond 5 meters depth.
Our findings suggest that sophisticated asymptotic analysis provides a more
accurate framework for capturing \gls{SA} model uncertainty nuances, offering
practical advantages in computational feasibility and interpretability.
While \cite{guillaume2023determination}'s method constrains inflated uncertainty
estimates from expected Fisher information, our comprehensive evaluation of
statistical tests demonstrates enhanced capacity to
capture inherent \gls{SA} model complexities and asymmetries.

Hyperspectral data emerges as more conducive for first-order statistics,
exhibiting favorable outcomes that underscore variance-covariance framework
adequacy. Conversely, multispectral data presents challenging landscapes for
first-order asymptotic theory, with notable concerns including \gls{WCD}
uncertainty asymmetry, practical identifiability issues, and sensitivity
to local optimizer performance. However, the likelihood ratio test accommodates
asymmetrical \gls{CI}—particularly relevant for multispectral
nonlinear \gls{SA} models. The identified behavioral patterns of the profile likelihood can
be related through Fisher information.
The ideal symmetric behavior observed in hyperspectral clear water scenarios
corresponds to cases with high Fisher information content, where the information
matrix is well-conditioned with large eigenvalues. Conversely, the practical 
identifiability limitations emerge when Fisher 
information on the \gls{WCD} parameter becomes critically low, 
often indicated by near-singular information
matrices with high condition numbers. This theoretical connection 
provides a quantitative criteria for predicting when traditional
Wald tests may fail and likelihood ratio approaches become necessary.

The benefits of \gls{MLE} refinement through profiling extend beyond our
controlled experiments initiated from true parameters. In practical scenarios,
reliance on global optimizers may be inefficient due to computational constraints
from vast spectral data volumes. Parameter estimation quality often depends
on optimizer choice and initial conditions \cite{Garcia2014}, frequently
resulting in suboptimal inference conditions. Our results indicate that
common local minima problems in \gls{SA} models can be effectively mitigated
through profiling, enhancing \gls{MLE} accuracy while enabling precise
uncertainty evaluation with fewer constraints compared to computationally
demanding methods like bootstrapping.

Our analysis reveals distinct performance patterns between sensor types 
that have important practical implications. While hyperspectral data enables 
reliable first-order statistical inference with rejection rates aligning 
with nominal values across diverse environmental conditions, multispectral 
applications face considerable challenges that emerge even in shallow clear waters.
Additionally, environmental optical complexity emerges as a critical determinant of uncertainty
quantification reliability. The systematic degradation from clear water conditions
to turbid environments reflects fundamental physical
limitations where multiple scattering and reduced light penetration limit 
information content regardless of sensor type. These environmental constraints
create depth-dependent uncertainty patterns that vary significantly across
coastal regions, suggesting that operational surveys must carefully 
consider local conditions when establishing protocols and interpreting uncertainty bounds.
The likelihood ratio test proves most robust across all conditions, maintaining 
reliable performance even when traditional variance-covariance approaches fail.

The asymmetric \gls{CI} reflect fundamental nonlinearities 
in \gls{SA} models and represent a feature, not a limitation. When spectral data
constrains minimum depth but not maximum depth, likelihood ratio test
correctly preserve this asymmetric information structure rather than 
forcing artificial symmetry. Even when upper bounds become unidentifiable,
reliable lower bounds provide critical minimum depth estimates for 
navigation safety. This asymmetric robustness reflects genuine 
physical limitations in light penetration rather than statistical inadequacy.

Profiling the likelihood requires additional optimization iterations with 
fixed interest parameters, representing a manageable computational overhead. 
This modest investment yields substantial gains in \gls{MLE} accuracy as well
as likelihood ratio confidence interval reliability, making it a worthwhile 
trade-off for operational applications prioritizing uncertainty quantification 
quality over processing speed.

Our controlled simulation framework, while essential for isolating statistical
effects, represents idealized conditions with perfect atmospheric correction
and sensor calibration. Real-world applications face additional uncertainties
from atmospheric correction, sensor degradation, and calibration drift.
The Gaussian error assumption may not capture complex error structures in
real spectral data, including systematic biases from atmospheric and adjacency
effects. Future work should investigate robust inference methods accommodating 
non-Gaussian error distributions and spatial correlation structures.
Our single-pixel analysis represents a challenging statistical
scenario but operational mapping often benefits from spatial context that
could improve uncertainty quantification.

Although Wald, score, and likelihood ratio tests are asymptotically first-order 
equivalent~\cite{BarndorffNielsenCox1989}, their finite-sample behavior differs substantially in nonlinear 
ill-conditioned inverse problems. In multispectral bathymetric retrievals, 
the effective sample size $n_\text{eff} \approx 2$---defined as the number of spectral 
bands minus the number of estimated parameters plus one, here 7 bands and $p=5$ 
parameters---creates a regime where $n_\text{eff}/p \approx 0.4$, far from the 
$n_\text{eff}/p \gg 1$ requirement of large-sample asymptotic theory~\cite{BarndorffNielsenCox1989}.
Limited spectral resolution causes near-collinearity in Jacobian 
columns when parameters produce similar spectral signatures, 
producing ill-conditioned Fisher information matrices.
Expected Fisher information provides a local quadratic
characterization of the likelihood surface at the \gls{MLE}; by contrast,
the likelihood ratio test evaluates likelihood differences along the 
full profile path, transcending this local approximation. As noted by 
Barndorff-Nielsen and Cox~\cite{BarndorffNielsenCox1989}, this property 
is peculiar to the log-likelihood ratio statistic: no equally simple 
general result holds for other first-order test statistics, which explains
its superior chi-squared approximation in higher-order asymptotics relevant
to small-sample regimes. Furthermore, in turbid water scenarios, strong correlation
among water-column parameters ($C_\text{chl}$, $C_\text{cdom}$, $C_\text{tsm}$) 
can overestimate the Fisher information, producing overly 
narrow---conservative---confidence intervals with high rejection rates,
as visible in Tables~\ref{tab:rej_rate_hys} and~\ref{tab:rej_rate_sen} for TWBB/TWDB 
at intermediate depths.

Profile likelihood provides practical advantages over alternative 
uncertainty methods for per-pixel bathymetric mapping. 
Bootstrapping~\cite{hedley2010propagating} requires repeated full 
inversions per pixel (e.g., 20-100 perturbations), making it 
computationally expensive. \cite{Garcia2015} demonstrated that this 
approach can suffer from optimization convergence issues when using 
standard local optimization algorithms, particularly in challenging 
conditions. 
Crucially, Monte-Carlo bootstrap estimates the average uncertainty 
expected across repeated experiments, not the conditional 
uncertainty for the specific observation at hand; since the \gls{MLE} is not 
a sufficient statistic in nonlinear models, these two quantities need 
not coincide, and it is presently unknown whether they differ significantly
in this context. Variance-covariance propagation~\cite{GUM,jay} 
via expected Fisher information is computationally efficient but produces
overly liberal confidence intervals in multispectral applications due
to the small-sample effects described above. Profile likelihood 
addresses both limitations: constrained optimizations during profiling 
are performed over $p-1$ parameters (with \gls{WCD} fixed), making each step 
intrinsically faster and better conditioned than the full $p$-parameter inversion;
the likelihood ratio statistic thereby provides asymmetric, observation-specific 
confidence intervals without the convergence failures of bootstrap.

While simulations are essential for rigorous coverage probability evaluation, 
they offer three specific advantages that real data cannot provide:
(1) the true value of all model parameters 
($z$, $C_\text{chl}$, $C_\text{cdom}$, $C_\text{tsm}$, $\alpha_m$) is
known exactly---not just \gls{WCD}; (2) the statistical inference performance
can be isolated from atmospheric correction and calibration confounders; 
and (3) our results represent lower bounds on real-world uncertainties,
since the perfect model assumption excludes bottom heterogeneity,
3D radiative transfer effects, and spectral variability of endmembers
that would further increase uncertainties in real data. 
Extending the study to real satellite data with concurrent 
acoustic ground truth would be a meaningful next step but would substantially
extend the scope of this methodological paper.

\section{Conclusion}

This paper's investigation into likelihood-based inference for bathymetric 
uncertainties within \gls{SA} models highlights the intricate balance between
theoretical precision and the pragmatic challenges encountered in real-world
environmental monitoring. Through rigorous examination of first-order inference, 
profiling refinement, and the distinct performance across hyperspectral 
and multispectral data, this study underscores the complexity of accurately
estimating \gls{WCD} and its associated uncertainties.

The application of \gls{MLE} inference through profiling emerged as a 
crucial technique, enhancing the robustness of parameter estimation under
the diverse conditions presented by \gls{SA} models. This methodology, proven effective 
across both idealized and challenging initial conditions, offers a pivotal 
advancement in the optimization landscape, particularly valuable in
the complex environments typical of bathymetric mapping where multiple
nuisance factors may be involved.

This study demonstrates that uncertainty
 quantification in \gls{SA} bathymetric
products must account for both sensor capabilities and environmental complexity.
For practitioners working with hyperspectral data in clear water conditions, 
traditional variance-covariance approaches (equivalent to expected Fisher 
information Wald tests) remain practical and computationally efficient, 
providing reliable symmetric confidence intervals that align well with 
established remote sensing uncertainty protocols. However, these methods 
deteriorate rapidly in multispectral applications and complex water environments. 

Among the evaluated tests, likelihood ratio test demonstrated
robustness in challenging scenarios, providing operationally 
useful asymmetric uncertainty bounds that match the information content in 
spectral data. Particularly, it outperformed the other tests in its ability to maintain 
reliable lower bound estimates even when upper bounds become unidentifiable.
The likelihood ratio test's asymmetric robustness represents a 
paradigm shift toward uncertainty quantification that honors the information structure
of spectral measurements rather than imposing unfounded statistical symmetry.
Future research should focus on developing adaptive uncertainty frameworks
and establishing operational protocols for
method selection based on spectral data 
characteristics and environmental conditions.
Future directions also include real-data validation with concurrent acoustic
 ground truth and investigation of robust inference methods for non-Gaussian error structures.

In conclusion, this study advocates for a methodical and cautious approach 
to the application of statistical inference within \gls{SA} models, emphasizing 
the need for further advancements to accurately quantify uncertainties 
across the spectrum of environmental conditions
encountered in operational coastal monitoring.

\appendix

\subsection{Variance-Covariance and Profile Likelihood Equivalence{\label{app:var_covar_equiv_mle}}}

For non-linear models $\mathbf{y} = \boldsymbol{\eta}(\boldsymbol{\theta}) + \boldsymbol{\epsilon}$ 
with $\boldsymbol{\epsilon} \sim \mathcal{N}(0, \boldsymbol{\Sigma})$,
linearization around $\boldsymbol{\hat{\theta}}$ yields \cite{Taylor1997}:

\begin{equation}
    \boldsymbol{\eta}(\boldsymbol{\theta}) \approx \boldsymbol{\eta}(\boldsymbol{\hat{\theta}}) + \mathbf{J}(\boldsymbol{\theta} - \boldsymbol{\hat{\theta}})
\end{equation}
where $\mathbf{J} = \frac{\partial \boldsymbol{\eta}}{\partial \boldsymbol{\theta}}$
is the Jacobian matrix.

The variance-covariance matrix becomes:
\begin{equation}
    \mathbf{\Sigma}_{\boldsymbol{\hat{\theta}}} = (\mathbf{J}^T \boldsymbol{\Sigma}^{-1} \mathbf{J})^{-1} = \boldsymbol{i}(\boldsymbol{\hat{\theta}})^{-1}
\end{equation}

For a parameter of interest $\psi$ within
$\boldsymbol{\theta} = (\psi, \boldsymbol{\omega})$, 
partitioning the Fisher information matrix \cite{severini}:
\begin{equation}
    \boldsymbol{i}(\boldsymbol{\theta}) = \begin{pmatrix}
        i_{\psi\psi} & \boldsymbol{i}_{\psi\boldsymbol{\omega}} \\
        \boldsymbol{i}_{\boldsymbol{\omega}\psi} & \boldsymbol{i}_{\boldsymbol{\omega\omega}}
    \end{pmatrix}
\end{equation}

The profile Fisher information for $\psi$ is:
\begin{equation}
    i_p(\psi) = i_{\psi\psi} - \boldsymbol{i}_{\psi\boldsymbol{\omega}} \boldsymbol{i}_{\boldsymbol{\omega\omega}}^{-1} \boldsymbol{i}_{\boldsymbol{\omega}\psi}
\end{equation}

This equals the inverse of the $(\psi,\psi)$ diagonal element
of $\boldsymbol{i}(\boldsymbol{\theta})^{-1}$, establishing 
equivalence between the expected Fisher information Wald test
and variance-covariance propagation for the 
parameter of interest under Gaussian errors.

\subsection{IOP modeling{\label{app:iops}}}

Here, we elaborate on the equations used to model the \gls{IOP}s of the water, specifically
the absorption \( \boldsymbol{a} \) and backscattering \( \boldsymbol{b_b} \) coefficients.

We consider the \gls{IOP}s of the water, specifically  for modeling the absorption \( \boldsymbol{a} \) and backscattering \( \boldsymbol{b_b} \) coefficients. 
In coastal waters, light is absorbed notably by water itself \( \boldsymbol{a}_{\text{w}} \), phytoplankton \( \boldsymbol{a}_{\phi} \), Colored Dissolved Organic Matter
(\gls{CDOM}) \( \boldsymbol{a}_{\text{cdom}} \) and the absorption $\boldsymbol{a}_{\text{nap}}$ due to suspended inorganic matter. 
These absorption contributions are modeled as follows (units in \(\text{m}^{-1}\)):
\begin{align}
    \boldsymbol{a}_{\phi
    }(\text{Chl}, \lambda) &= \boldsymbol{\gamma}_{\phi}(\lambda) \cdot \text{Chl}^{\boldsymbol{\nu}_{\phi}(\lambda)}   \\
    \boldsymbol{a}_{\text{cdom}}(\text{CDOM} ,\lambda) &= \text{CDOM} \cdot \exp\left(-\beta_{\text{cdom}} \cdot (\lambda - 443)\right) \\
    \boldsymbol{a}_{\text{nap}}(\text{TSM} , \lambda) &= \text{TSM} \cdot \gamma_{\text{nap}} \cdot \exp\left(-\beta_{\text{nap}} \cdot (\lambda - 443)\right) 
\end{align}
$\boldsymbol{a}_w$ is provided in \cite{Pope97}, $\boldsymbol{a}_\phi$ is modeled with the empirical spectra
$\boldsymbol{\gamma}_{\phi}(\lambda),\boldsymbol{\nu}_{\phi}(\lambda)$ derived from \cite{bricaud_chl} whereas the 
empirical values $\beta_{\text{cdom}}$, $\gamma_{nap}$ abd $\beta_{\text{nap}}$ are set to .0176, 0.036, 0.0123 respectively, as
average values for coastal waters as proposed in \cite{bricaud}.

Contributions to the backscattering \( \boldsymbol{b_{b}} \) originate mainly from water \( \boldsymbol{b_{w}} \) 
and suspended particulate matter \( \boldsymbol{b_{\text{p}}} \). 
\begin{equation}
    \boldsymbol{b_b} = \boldsymbol{{b_{w}} + \boldsymbol{b_{\text{p}}}}
\end{equation}
\begin{align}
    \boldsymbol{b_{\text{p}}}(\text{TSM},\lambda) &=   \text{TSM} \cdot \gamma_{\text{bp}} \cdot \left(\frac{\lambda}{555.0}\right)^{-\nu_{\text{bp}}} , \\
    \boldsymbol{b_{\text{w}}}(\lambda) &=  \gamma_{\text{bw}} \cdot \left(\frac{\lambda}{500.0}\right)^{\nu_{\text{bw}}}.
\end{align}
The spectral power coefficient for suspended particulate matter backscattering $\nu_{\text{bp}}$
 varies globally and seasonally within the range 0.5,4.5 depending on the empirical algorithm 
 \cite{bpglobal}. We set these empirical values  at $\gamma_{\text{bp}} = 0.0077$, $\nu_{\text{bp}} = -0.2$ from \cite{lee1999hyperspectral}. 
For water backscattering, values were fixed at at$\gamma_{\text{bw}} = 0.0035$, $\nu_{\text{bw}} = -4.32$ from \cite{Morel}. 
The water column \gls{IOP}'s parameter is then \( \boldsymbol{\omega}_c = (C_{\text{chl}}, C_{\text{cdom}}, C_{\text{tsm}}) \), where $C_{\text{Chl}}$ is in \( \text{mg m}^{-3} \), 
$C_{\text{cdom}}$ in \( \text{m}^{-1} \), $C_{\text{tsm}}$ in \( \text{g m}^{-3} \).


\bibliographystyle{IEEEtran}
\bibliography{references}  

\begin{thebibliography}{10}
\providecommand{\url}[1]{#1}
\csname url@samestyle\endcsname
\providecommand{\newblock}{\relax}
\providecommand{\bibinfo}[2]{#2}
\providecommand{\BIBentrySTDinterwordspacing}{\spaceskip=0pt\relax}
\providecommand{\BIBentryALTinterwordstretchfactor}{4}
\providecommand{\BIBentryALTinterwordspacing}{\spaceskip=\fontdimen2\font plus
\BIBentryALTinterwordstretchfactor\fontdimen3\font minus
  \fontdimen4\font\relax}
\providecommand{\BIBforeignlanguage}[2]{{%
\expandafter\ifx\csname l@#1\endcsname\relax
\typeout{** WARNING: IEEEtran.bst: No hyphenation pattern has been}%
\typeout{** loaded for the language `#1'. Using the pattern for}%
\typeout{** the default language instead.}%
\else
\language=\csname l@#1\endcsname
\fi
#2}}
\providecommand{\BIBdecl}{\relax}
\BIBdecl

\bibitem{vazquez}
M.~L. Martínez, A.~Intralawan, G.~Vázquez, O.~Pérez-Maqueo, P.~Sutton, and
  R.~Landgrave, ``The coasts of our world: Ecological, economic and social
  importance,'' \emph{Ecological Economics}, vol.~63, pp. 254--272, 8 2007.

\bibitem{Singh2021}
G.~G. Singh, R.~S. Cottrell, T.~D. Eddy, and A.~M. Cisneros-Montemayor,
  ``Governing the land-sea interface to achieve sustainable coastal
  development,'' \emph{Frontiers in Marine Science}, vol.~8, Jul. 2021, art.
  no. 709947.

\bibitem{Orton2020}
\BIBentryALTinterwordspacing
P.~Orton, E.~Sanderson, S.~Talke, M.~Giampieri, and K.~Macmanus, ``Storm tide
  amplification and habitat changes due to urbanization of a lagoonal
  estuary,'' \emph{Natural Hazards and Earth System Sciences}, 2020. [Online].
  Available:
  \url{https://nhess.copernicus.org/articles/20/2415/2020/nhess-20-2415-2020.pdf}
\BIBentrySTDinterwordspacing

\bibitem{Iryanthony2019}
\BIBentryALTinterwordspacing
S.~B. Iryanthony, H.~Haeruddin, M.~Helmi, and P.~A. Macklin, ``Utilizing
  landsat satellite data (1990-2018) to detect water inundation for the
  management of human settlements in coastal zones,'' \emph{Indian Journal of
  Science and Technology}, 2019. [Online]. Available:
  \url{https://indjst.org/download-article.php?Article_Unique_Id=INDJST2657&Full_Text_Pdf_Download=True}
\BIBentrySTDinterwordspacing

\bibitem{Pasquali2021}
D.~Pasquali and A.~Marucci, ``The effects of urban and economic development on
  coastal zone management,'' \emph{Sustainability (Switzerland)}, vol.~13, 6
  2021.

\bibitem{hydrobook}
\BIBentryALTinterwordspacing
M.~G. Cain, G.~B. Rossi, M.~V. Veghel, and K.-Y. Jhang, \emph{Measurement for
  the Sea}, ser. Springer Series in Measurement Science and Technology.\hskip
  1em plus 0.5em minus 0.4em\relax Cham, Switzerland: Springer International
  Publishing, 2022. [Online]. Available:
  \url{https://link.springer.com/book/10.1007/978-3-030-82558-2}
\BIBentrySTDinterwordspacing

\bibitem{bernardis2023}
\BIBentryALTinterwordspacing
M.~Bernardis, R.~Nardini, L.~Apicella, M.~Demarte, M.~Guideri, B.~Federici,
  A.~Quarati, and M.~De~Martino, ``Use of icesat-2 and sentinel-2 open data for
  the derivation of bathymetry in shallow waters: Case studies in sardinia and
  in the venice lagoon,'' \emph{Remote Sensing}, vol.~15, no.~11, 2023.
  [Online]. Available: \url{https://www.mdpi.com/2072-4292/15/11/2944}
\BIBentrySTDinterwordspacing

\bibitem{Ashphaq2021}
M.~Ashphaq, P.~K. Srivastava, and D.~Mitra, ``Review of near-shore satellite
  derived bathymetry: Classification and account of five decades of coastal
  bathymetry research,'' \emph{Journal of Ocean Engineering and Science},
  vol.~6, pp. 340--359, 12 2021.

\bibitem{lyzenga}
D.~Lyzenga, ``Passive remote sensing techniques for mapping water depth and
  bottom features,'' \emph{Applied Optics}, vol. 379, February 1978.

\bibitem{lee98}
Z.~Lee, K.~L. Carder, C.~D. Mobley, R.~G. Steward, and J.~S. Patch,
  ``Hyperspectral remote sensing for shallow waters. {A} semi-analytical
  model,'' \emph{Applied Optics}, vol.~37, no.~27, pp. 6329--6338, Sep. 1998.

\bibitem{lee1999hyperspectral}
------, ``Hyperspectral remote sensing for shallow waters: 2. deriving bottom
  depths and water properties by optimization,'' \emph{Applied Optics},
  vol.~38, no.~18, pp. 3831--3843, 1999.

\bibitem{stumpf}
R.~Stumpf, K.~Holderied, and M.~Sinclair, ``Determination of water depth with
  high-resolution satellite imagery over variable bottom types,''
  \emph{Limnology and Oceanography}, vol.~48, pp. 547,557, 2003.

\bibitem{Manessa2016}
M.~D.~M. Manessa, A.~Kanno, M.~Sekine, M.~Haidar, K.~Yamamoto, T.~Imai, and
  T.~Higuchi, ``Satellite-derived bathymetry using random forest algorithm and
  worldview-2 imagery,'' \emph{Geoplanning: Journal of Geomatics and Planning},
  vol.~3, p. 117, 10 2016.

\bibitem{Benshila2022}
R.~Benshila, Y.~E. Bennioui, G.~Thoumyre, R.~Almar, E.~W. Bergsma, M.~A. Najar,
  J.~M. Delvit, and D.~G. Wilson, ``Coastal bathymetry estimation from
  sentinel-2 satellite imagery: Comparing deep learning and physics-based
  approaches,'' \emph{Remote Sensing}, vol.~14, 3 2022.

\bibitem{lubac2022}
\BIBentryALTinterwordspacing
B.~Lubac, O.~Burvingt, A.~N. Lerma, and N.~Sénéchal, ``Performance and
  uncertainty of satellite-derived bathymetry empirical approaches in an
  energetic coastal environment,'' \emph{Remote Sensing}, vol.~14, no.~10, p.
  2350, 2022. [Online]. Available:
  \url{https://www.mdpi.com/2072-4292/14/10/2350/pdf?version=1652360089}
\BIBentrySTDinterwordspacing

\bibitem{Garcia2015}
R.~A. Garcia, ``Uncertainty in hyperspectral remote sensing: Analysis of the
  potential and limitation of shallow water bathymetry and benthic
  classification,'' Ph.D. dissertation, Dept. Phys. Astron., Curtin Univ.,
  Perth, WA, Australia, 2015.

\bibitem{jay}
S.~Jay, M.~Guillaume, M.~Chami, Y.~D. Audrey~Minghelli, B.~Lafrance, and
  V.~Serfaty, ``Predicting minimum uncertainties in the inversion of ocean
  color geophysical parameters based on cramer-rao bounds,'' \emph{OPTICS
  EXPRESS}, vol.~26, January 2018.

\bibitem{jaybottom}
J.~Sylvain, M.~Guillaume, A.~Minghelli, Y.~Deville, M.~Chami, L.~Bruno, and
  V.~Serfaty, ``Hyperspectral remote sensing of shallow waters: Considering
  environmental noise and bottom intra-class variability for modeling and
  inversion of water reflectance,'' \emph{Remote Sensing of Environment}, vol.
  200, pp. 352,367, October 2017.

\bibitem{1er}
M.~A. Ghannami, G.~Sicot, M.~Lennon, S.~Loyer, and N.~Thomas, ``Likelihood
  ratio statistic for inferring the uncertainty of satellite derived
  bathymetry,'' in \emph{2021 11th Workshop on Hyperspectral Imaging and Signal
  Processing: Evolution in Remote Sensing (WHISPERS)}, 2021, pp. 1--5.

\bibitem{Gao2009}
B.-C. Gao, M.~J. Montes, C.~O. Davis, and A.~F. Goetz, ``Atmospheric correction
  algorithms for hyperspectral remote sensing data of land and ocean,''
  \emph{Remote Sensing of Environment}, vol. 113, pp. S17--S24, 2009.

\bibitem{hedley}
J.~Hedley, C.~Roelfsemab, B.~Koetzc, and S.~Phinn, ``Capability of the sentinel
  2 mission for tropical coral reef mapping and coral bleaching detection,''
  \emph{Remote Sensing of Environment}, February 2012.

\bibitem{mckinna2019approach}
L.~I.~W. McKinna, I.~Cetini\'{c}, A.~P. Chase, and P.~J. Werdell, ``Approach
  for propagating radiometric data uncertainties through {NASA} ocean color
  algorithms,'' \emph{Frontiers in Earth Science}, vol.~7, Jul. 2019, art. no.
  176.

\bibitem{guillaume2023determination}
M.~Guillaume, A.~Minghelli, M.~Chami, and M.~Lei, ``Determination of bayesian
  cram\'{e}r--rao bounds for estimating uncertainties in the bio-optical
  properties of the water column, the seabed depth and composition in a coastal
  environment,'' \emph{Remote Sensing}, vol.~15, no.~9, p. 2242, 2023.

\bibitem{Hu2013}
Z.~L. Chuanmin~Hu, Lian~Feng, ``Uncertainties of seawifs, modis remote sensing
  reflectance: Implications from clear water measurements,'' \emph{Remote
  Sensing of Environment}, vol. 133, pp. 168--182, 6 2013.

\bibitem{melin2015}
\BIBentryALTinterwordspacing
F.~Mélin, G.~Sclep, T.~Jackson, and S.~Sathyendranath, ``Uncertainty estimates
  of remote sensing reflectance derived from comparison of ocean color
  satellite data sets,'' \emph{Remote Sensing of Environment}, vol. 177, pp.
  107--124, 2016. [Online]. Available:
  \url{https://www.sciencedirect.com/science/article/pii/S0034425716300426}
\BIBentrySTDinterwordspacing

\bibitem{Gordon1997}
H.~R. Gordon, ``Atmospheric correction of ocean color imagery in the earth
  observing system era,'' \emph{Journal of Geophysical Research}, vol. 102, no.
  D14, pp. 17,081--17,106, 1997.

\bibitem{Wettle2004}
M.~Wettle, V.~E. Brando, and A.~G. Dekker, ``A methodology for retrieval of
  environmental noise equivalent spectra applied to four hyperion scenes of the
  same tropical coral reef,'' \emph{Remote Sensing of Environment}, vol.~93,
  pp. 188--197, 10 2004.

\bibitem{Sagar2014}
S.~Sagar, V.~Brando, and M.~Sambridge, ``Noise estimation of remote sensing
  reflectance using a segmentation approach suitable for optically shallow
  waters,'' \emph{IEEE Transactions on Geoscience and Remote Sensing}, vol.~52,
  pp. 7504--7512, 2014.

\bibitem{Garcia2014}
R.~A. Garcia, L.~I. McKinna, J.~D. Hedley, and P.~R. Fearns, ``Improving the
  optimization solution for a semi-analytical shallow water inversion model in
  the presence of spectrally correlated noise,'' \emph{Limnology and
  Oceanography: Methods}, vol.~12, pp. 651--669, 2014.

\bibitem{Jia2018}
J.~Guorui, X.~Qian, and Z.~Huijie, ``Uncertainty analysis for surface
  reflectance retrieved from hyperspectral remote sensing image using empirical
  line method,'' in \emph{2018 9th Workshop on Hyperspectral Image and Signal
  Processing: Evolution in Remote Sensing (WHISPERS)}, 2018, pp. 1--5.

\bibitem{nicolas}
N.~Thomas, M.~Lennon, C.~Danilo, G.~Sicot, M.~A. Ghannami, and S.~Loyer, ``A
  method for propagating uncertainties of the top of atmosphere sentinel-2
  measurements to bottom of atmosphere reflectance for aquatic applications,''
  in \emph{2021 11th Workshop on Hyperspectral Imaging and Signal Processing:
  Evolution in Remote Sensing (WHISPERS)}, 2021, pp. 1--5.

\bibitem{pachniak2023quantifying}
E.~Pachniak, Y.~Fan, W.~Li, and K.~Stamnes, ``Quantifying uncertainties in
  oc-smart ocean color retrievals: A bayesian inversion algorithm,''
  \emph{Algorithms}, vol.~16, no.~6, p. 301, 2023.

\bibitem{hedley2010propagating}
J.~Hedley, C.~Roelfsema, and S.~Phinn, ``Propagating uncertainty through a
  shallow water mapping algorithm based on radiative transfer model
  inversion,'' in \emph{Proceedings of Ocean Optics XX}, vol.~12, Anchorage,
  AK, USA, 2010, pp. 24--51.

\bibitem{petit}
\BIBentryALTinterwordspacing
T.~Petit, T.~Bajjouk, P.~Mouquet, S.~Rochette, B.~Vozel, and C.~Delacourt,
  ``{Hyperspectral remote sensing of coral reefs by semi-analytical model
  inversion -- Comparison of different inversion setups},'' \emph{{Remote
  Sensing of Environment}}, vol. 190, pp. 348--365, 2017. [Online]. Available:
  \url{https://univ-rennes.hal.science/hal-01484526}
\BIBentrySTDinterwordspacing

\bibitem{GUM}
{Joint Committee for Guides in Metrology}, \emph{Evaluation of Measurement Data
  -- Guide to the Expression of Uncertainty in Measurement}, 1st~ed., Bureau
  International des Poids et Mesures, Sèvres, France, 2008, jCGM 100:2008.

\bibitem{lee2010uncertainties}
Z.~Lee, R.~Arnone, C.~Hu, P.~J. Werdell, and B.~Lubac, ``Uncertainties of
  optical parameters and their propagations in an analytical ocean color
  inversion algorithm,'' \emph{Applied Optics}, vol.~49, no.~3, pp. 369--381,
  2010.

\bibitem{maritorena2004consistent}
\BIBentryALTinterwordspacing
S.~Maritorena and D.~A. Siegel, ``Consistent merging of satellite ocean color
  data sets using a bio-optical model,'' \emph{Remote Sensing of Environment},
  vol.~94, no.~4, pp. 429--440, 2005. [Online]. Available:
  \url{https://www.sciencedirect.com/science/article/pii/S0034425704002834}
\BIBentrySTDinterwordspacing

\bibitem{salama2011ensemble}
M.~S. Salama, F.~Melin, and R.~Van~der Velde, ``Ensemble uncertainty of
  inherent optical properties,'' \emph{Optical Society of America}, 2011.

\bibitem{Sydor2003}
M.~Sydor, R.~Gould, R.~Arnone, V.~Haltrin, and W.~Goode, ``Uniqueness in remote
  sensing of the inherent optical properties of ocean water,'' \emph{Applied
  optics}, vol.~43, pp. 2156--62, 04 2004.

\bibitem{DefoinPlatel2007}
M.~Defoin-Platel and M.~Chami, ``How ambiguous is the inverse problem of ocean
  color in coastal waters?'' \emph{Journal of Geophysical Research}, vol. 112,
  p. C03004, 2007.

\bibitem{Garcia2018}
R.~A. Garcia, Z.~Lee, and E.~J. Hochberg, ``Hyperspectral shallow-water remote
  sensing with an enhanced benthic classifier,'' \emph{Remote Sensing},
  vol.~10, no.~1, p. 147, 2018.

\bibitem{sicot}
G.~Sicot, M.~A. Ghannami, M.~Lennon, and S.~Loyer, ``Estimability study of the
  parameters of the semi-analytical lee model with hyperspectral data,''
  \emph{IEEE WHISPERS}, p. Unknown Pages, 2021.

\bibitem{Wieland2021}
F.-G. Wieland, A.~L. Hauber, M.~Rosenblatt, C.~Tönsing, and J.~Timmer, ``On
  structural and practical identifiability,'' \emph{Current Opinion in Systems
  Biology}, vol.~25, pp. 60--69, March 2021.

\bibitem{pronzato}
L.~Pronzato and A.~Pazman, \emph{Design of Experiments in Nonlinear
  Models}.\hskip 1em plus 0.5em minus 0.4em\relax Springer, 2013.

\bibitem{Lam2022}
R.~M. Nicholas N.~Lam, Paul D.~Docherty, ``Practical identifiability of
  parametrised models: A review of benefits and limitations of various
  approaches,'' \emph{Mathematics and Computers in Simulation}, vol. 199, pp.
  202--216, September 2022.

\bibitem{brazzal}
\BIBentryALTinterwordspacing
A.~R. Brazzale, A.~C. Davison, and N.~Reid, \emph{Applied Asymptotics: Case
  Studies in Small-Sample Statistics}.\hskip 1em plus 0.5em minus 0.4em\relax
  Cambridge: Cambridge University Press, 2007. [Online]. Available:
  \url{http://infoscience.epfl.ch/record/104219}
\BIBentrySTDinterwordspacing

\bibitem{Pope97}
\BIBentryALTinterwordspacing
R.~M. Pope and E.~S. Fry, ``Absorption spectrum (380--700 nm) of pure water.
  ii. integrating cavity measurements,'' \emph{Appl. Opt.}, vol.~36, no.~33,
  pp. 8710--8723, Nov 1997. [Online]. Available:
  \url{https://opg.optica.org/ao/abstract.cfm?URI=ao-36-33-8710}
\BIBentrySTDinterwordspacing

\bibitem{bricaud_chl}
A.~Bricaud, M.~Babin, A.~Morel, and H.~Claustre, ``Variability in the
  chlorophyll-specific absorption coefficients of natural phytoplankton:
  Analysis and parametrization,'' \emph{Journal of Geophysical Research}, vol.
  100, pp. 13,321--13,332, 07 1995.

\bibitem{bricaud}
M.~Babin, D.~Stramski, G.~M. Ferrari, H.~Claustre, A.~Bricaud, G.~Obolensky,
  and N.~Hoepffner, ``Variations in the light absorption coefficients of
  phytoplankton, nonalgal particles, and dissolved organic matter in coastal
  waters around {Europe},'' \emph{Journal of Geophysical Research: Oceans},
  vol. 108, no.~C7, Jul. 2003, art. no. 3211.

\bibitem{bpglobal}
\BIBentryALTinterwordspacing
X.~Yu, Z.~Lee, and W.~Lai, ``Global distribution of the spectral power
  coefficient of particulate backscattering coefficient obtained by a neural
  network scheme,'' \emph{Remote Sensing of Environment}, vol. 296, p. 113750,
  2023. [Online]. Available:
  \url{https://www.sciencedirect.com/science/article/pii/S0034425723003012}
\BIBentrySTDinterwordspacing

\bibitem{Morel}
A.~Morel, ``Optical properties of pure water and pure sea water,'' in
  \emph{Optical Aspects of Oceanography}, N.~G. Jerlov and E.~Steemann~Nielsen,
  Eds.\hskip 1em plus 0.5em minus 0.4em\relax New York, NY, USA: Academic
  Press, 1974, pp. 1--24.

\bibitem{Tocquet2001}
\BIBentryALTinterwordspacing
A.~S. Tocquet, ``Likelihood based inference in non-linear regression models
  using the p* and r* approach,'' \emph{Scandinavian Journal of Statistics},
  vol.~28, no.~3, pp. 429--443, 2001. [Online]. Available:
  \url{https://www.jstor.org/stable/4616670}
\BIBentrySTDinterwordspacing

\bibitem{ISO11664-1}
{International Commission on Illumination (CIE)}, \emph{Colorimetry — Part 1:
  CIE Standard Colorimetric Observers}, Std. ISO/CIE 11\,664-1:2019, 2019.

\bibitem{trf}
\BIBentryALTinterwordspacing
M.~A. Branch, T.~F. Coleman, and Y.~Li, ``A subspace, interior, and conjugate
  gradient method for large-scale bound-constrained minimization problems,''
  \emph{SIAM Journal on Scientific Computing}, vol.~21, no.~1, pp. 1--23, 1999.
  [Online]. Available: \url{https://doi.org/10.1137/S1064827595289108}
\BIBentrySTDinterwordspacing

\bibitem{BarndorffNielsenCox1989}
O.~E. Barndorff-Nielsen and D.~R. Cox, \emph{Asymptotic Techniques for Use in
  Statistics}.\hskip 1em plus 0.5em minus 0.4em\relax London: Chapman and Hall,
  1989, chapter 4 discusses profile likelihood and higher-order approximations
  for likelihood ratio tests.

\bibitem{Taylor1997}
J.~R. Taylor, \emph{An Introduction to Error Analysis: The Study of
  Uncertainties in Physical Measurements}, 2nd~ed.\hskip 1em plus 0.5em minus
  0.4em\relax Sausalito, CA: University Science Books, 1997.

\bibitem{severini}
T.~A. Severini, \emph{Likelihood Methods in Statistics}.\hskip 1em plus 0.5em
  minus 0.4em\relax Oxford, U.K.: Oxford Univ. Press, 2000.

\end{thebibliography}

\section{Author Biographies}

\begin{IEEEbiographynophoto}{Mohamed Ali Ghannami}
is affiliated with the National Institute
of Advanced Technologies, (Lab-STICC, Brest, France). 
His research focuses on spectral bathymetry, uncertainty
quantification in semi-analytical radiative transfer models, 
and statistical inference for optical remote sensing of shallow waters. 
He has published work on likelihood-based
triangulation methods for estimating through-water depth uncertainties.

\end{IEEEbiographynophoto}

\begin{IEEEbiographynophoto}{Guillaume Sicot}
is also with Lab-STICC, National
Institute of Advanced Technologies of Brittany, Brest, France. 
He has published extensively in coastal and aquatic remote sensing,
especially in radiative transfer modeling, parameter estimation, 
and uncertainty analysis for bathymetric inversion and 
optical water column modeling. 

\end{IEEEbiographynophoto}

\begin{IEEEbiographynophoto}{Isabelle Quidu}
is a Professor (Associate) with Lab-STICC UMR CNRS 6285 and HDR at 
the National Institute of Advanced Technologies, Brest, France. She received the Ph.D. degree 
in electronics from the University of Western Brittany, Brest, France, in 2001, and the Habilitation \`{a} Diriger des Recherches (HDR) from the same university in 2018.
Her expertise lies in
underwater imaging, sonar signal processing, 
and multi-sensor data fusion for marine applications.

\end{IEEEbiographynophoto}

\begin{IEEEbiographynophoto}{Sylvie Daniel}
is a Full Professor in the Department of
geomatic sciences at Université Laval, Québec, Canada. 
She earned her Ph.D. in signal processing and telecommunications
from Rennes I University, Rennes, France, in 1998. 
Her research interests include data acquisition (bathymetric and LiDAR point clouds, imagery), 
geospatial and hydrospatial data processing, artificial intelligence, 
3D modeling, data fusion, and uncertainty assesment. 
She is Senior Member of IEEE and member of the OES and GRSS societies.
\end{IEEEbiographynophoto}
\end{document}